\documentclass[onecolumn,draftcls]{IEEEtran}
\usepackage[utf8]{inputenc}
\usepackage{graphicx, graphics}
\usepackage{amsmath, amsthm, amssymb}
\usepackage{color}
\usepackage{cite}
\usepackage{booktabs}
\usepackage{multirow}
\usepackage{float}
\restylefloat{table}
\usepackage{makecell}
\usepackage{flushend}

\DeclareMathAlphabet\mathbfcal{OMS}{cmsy}{b}{n}
\theoremstyle{remark}
\newtheorem{remark}{Remark}

\def\TxRisChannel{\mathbf{G}}
\def\TxRisChannelvec{\mathbf{g}}
\def\RisRxChannel{\mathbf{H}}
\def\RisRxChannelvec{\mathbf{h}}
\def\TxRxChannel{\mathbf{H}_d}
\def\TxRxChannelvec{\mathbf{h}_d}

\def\CascadedChannel{\mathbf{H}_c}
\def\CascadedChannelvec{\mathbf{h}_c}
\def\CascadedChannelvecEst{\hat{\mathbf{h}}_c}
\def\RisPhaseMatrix{\mathbf{\Phi}}
\def\RisPhasevec{\boldsymbol{\psi}}
\def\inputvec{\mathbf{x}}
\def\outputvec{\mathbf{y}}
\def\noisevec{\mathbf{w}}
\def\diag{\text{diag}}
\def\vec{\text{vec}}
\def\identity{\mathbf{I}}
\def\bPsi{\boldsymbol{\Psi}}

\def\bA{\boldsymbol{A}}
\def\bB{\boldsymbol{B}}
\def\bLambda{\boldsymbol{\Lambda}}
\def\blambda{\boldsymbol{\lambda}}
\def\ostar{\circledast}
\def\Figwidth{3.5 in}
\def\TxBeamformer{\mathbf{f}}
\def\argmin{\text{argmin}}
\def\BibTeX{{\rm B\kern-.05em{\sc i\kern-.025em b}\kern-.08em
    T\kern-.1667em\lower.7ex\hbox{E}\kern-.125emX}}

\DeclareFontFamily{U}{mathx}{\hyphenchar\font45}
\DeclareFontShape{U}{mathx}{m}{n}{
      <5> <6> <7> <8> <9> <10>
      <10.95> <12> <14.4> <17.28> <20.74> <24.88>
      mathx10
      }{}
\DeclareSymbolFont{mathx}{U}{mathx}{m}{n}
\DeclareFontSubstitution{U}{mathx}{m}{n}
\DeclareMathAccent{\widecheck}{0}{mathx}{"71}
\DeclareMathAccent{\wideparen}{0}{mathx}{"75}

\begin{document}

\title{Channel Training \& Estimation for Reconfigurable Intelligent Surfaces: Exposition of Principles, Approaches, and Open Problems}
\author{Bharath Shamasundar, Negar Daryanavardan, and Aria Nosratinia\thanks{This work was supported in part by the National Science Foundation under Grants 1956213 and 2148211.}}

\maketitle

\begin{abstract}
Reconfigurable intelligent surfaces (RIS) are passive controllable arrays of small reflectors that direct electromagnetic energy towards or away from the target nodes, thereby allowing better management of signals and interference in a wireless network. 
The RIS has the potential for significantly improving the performance of wireless networks. Unfortunately, RIS also multiplies the number of Channel State Information (CSI) coefficients between the transmitter and receiver, which magnifies the challenges in estimating and communicating the channel state information. Furthermore, the simplicity and cost-effectiveness of the passive RIS also implies that the incoming links are not {\em locally} estimated at the RIS, and fresh pilots are not inserted into outgoing RIS links. This introduces {\em new} challenges for training and estimation of channel state information. The rapid growth of the literature on CSI acquisition in RIS-aided systems has been accompanied by variations in the underlying assumptions, models, and notation, which can obscure the similarities and differences of various techniques, and their relative merits.
This paper presents a comprehensive exposition of principles and approaches in RIS channel estimation. The basic ideas underlying each class of techniques are reduced to their simplest form under a unified model and notation, and various approaches within each class are discussed. 
Several open problems in this area are identified and highlighted.
\end{abstract}



\section{Introduction}
\label{sec:introduction}
The fifth generation (5G) wireless systems are being successfully deployed across the globe, achieving high data rates using large antenna arrays~\cite{7894280, 6798744}. 6G systems will aim for even higher data rates, lower latency, and better reliability, at low cost/complexity and high energy efficiency \cite{8766143,8869705}. This requires innovations in the physical layer technologies. 
The continuing migration to the millimeter wave (mmWave) spectrum, while improving data rates, has challenges including channel blockage and intermittent availability. Increasing the network density can solve some of these problems, but it injects more power into the network, increases the interference levels, escalates cost of deployment and operation, and raises scalability concerns. 

\begin{figure}[h]
    \centering
\includegraphics[width=3.5 in]{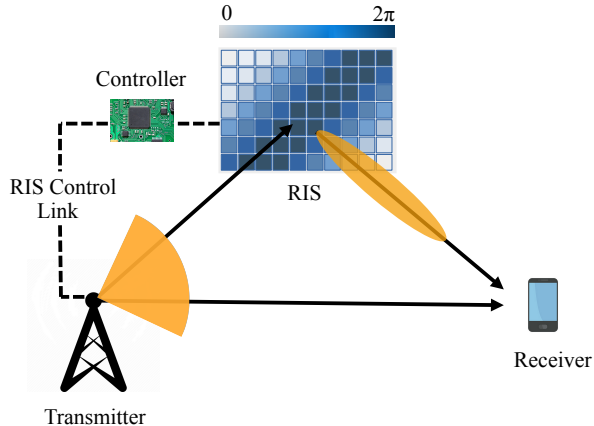}
    \vspace{-1 cm}
    \caption{Schematic representation of reconfigurable intelligent surface aided communication. }
    \label{fig:RIS-Schematic}
\end{figure}

Reconfigurable intelligent surfaces (RIS) are passive, controllable arrays of small reflectors that direct radio waves toward or away from a target node, enabling better management of signals and interference in wireless networks \cite{di2019smart, 9424177, 8796365, 8910627,9475160}. 
They are often interpreted as a mechanism that achieves software-defined control of the wireless propagation environment (Fig. 1). This approach has the potential to address several of the challenges mentioned above. Judiciously altering the channel characteristics in real-time as a part of the system operation, to achieve favorable propagation environment, is an attractive idea to handle challenging channel conditions. Unlike a relay, RIS does not inject more transmit power into the network, and its operation is independent of the details of PHY signaling other than operating frequency and bandwidth. The combination of lower power consumption and simpler construction makes it cheaper to build and operate, thus helping scalability.

RIS-induced channel coefficients must be estimated at the receiver for coherent communication, and shared with the transmitter (and RIS) for beamforming. The RIS channel estimation problem is distinct from the MIMO (multiple-input multiple-output) case because:  (1)~RIS is a two-hop channel observable only end-to-end, due to passivity of RIS, (2)~the number of channel gain coefficients is multiplied by the RIS size, making for a larger vector to be estimated, and (3)~in RIS channel estimation, training occurs through pilots plus passive RIS training states, and the latter is without a direct counterpart in traditional channel training and estimation.

\subsection{Contributions and Distinctions of the Present Work}

The present work provides a thorough exposition of the ideas underlying the rapidly expanding literature on RIS channel estimation, in a way that is both comprehensive and accessible to a wider audience. An up-to-date discussion of the available estimation techniques and related issues, ranging from classical least squares to the most recent artificial intelligence and machine learning (AI/ML)-based methods, is provided.

One of the contributions of the present work is to bring a unified notation and system model to the mathematical expression of various RIS channel estimation problems and algorithms. This goes beyond the enumeration of works in a conventional survey, and is greatly helpful in the interpretation and comparison of results in a rapidly expanding literature. Among other features, this work illuminates the commonality and differences between RIS channel estimation results, thus exposing synergies and facilitating the generation of new ideas.

In addition to the exposition of various estimation approaches, this paper also explores critical assumptions, explicit or implicit in RIS channel estimation, that are in need of further investigation and validation, e.g., channel reciprocity or perfect deactivation of RIS elements. Other important issues, such as the near field effect in very large RIS, or the dependence of reflection gains on phases, are also discussed. 

To put this paper in context and highlight its distinctions, we briefly review related works. In a concise letter, Wei {\em et al.}~\cite{9328501} explored sparsity, user correlation, and time scales for RIS channel estimation. Swindlehurst {\em et al.}~\cite{9328501} consider the identifiability of the models as a function of the pilots and RIS training states, and further consider special cases such as single-input single-output (SISO) and MIMO, availability or unavailability of a direct link, and narrowband vs. wideband. Noh {\em et al.}~\cite{9757755} concentrates on characteristics of RIS channels in terahertz (THz) and mmWave channels. Zheng {\em et al.}~\cite{9722893} provides a survey of RIS channel estimation that enumerates the problems and outcomes, but does not dwell on methodology or characterization of the methods. In comparison, the present work provides a simple yet sufficiently detailed exposition of ideas for a wider audience. Compared with the works mentioned above, the present work also addresses new dimensions, including: the impact of RIS channel estimation overhead on spectral efficiency, and the implications of optimizing spectral efficiency (through channel estimation constraints) on the size of RIS array. The present work also includes coverage of machine learning methods for RIS channel estimation. Also, as mentioned earlier, underlying assumptions with potential impact on practical implementations are also examined in the present work. 

This paper also presents several new interpretations and connections that have been unavailable in the literature thus far. Among them,  (a)~Section~\ref{sec:CommonRisBsChannel} provides an elegant explanation for available savings in multiuser RIS systems, leading to suggestions for future work.
(b)~Section~\ref{sec:statistical-csi} discusses the differences and similarities of techniques that infrequently update the RIS coefficients, compared with traditional MIMO statistical CSI methods, opening venues for future work, and (c)~Section~\ref{sec:OpportunisticRIS} clarifies for the first time that the so-called ``codebook methods'' for RIS actually have deep connections with opportunistic communication.

The organization of this paper is as follows. Section~\ref{sec:System-Model} presents RIS system and channel models. Section~\ref{sec:linear-estimation} formulates RIS channel estimation problem and presents discussions on pilots, training states, and linear estimation methods. Section~\ref{sec:SE-Training} presents the effect of training overhead and accuracy on the spectral efficiency of RIS, and discusses its impact on the choice of RIS dimensionality. Section~\ref{sec:sparse-estimation} presents the formulation of RIS channel estimation as a sparse recovery problem when operating in high frequency (mmWave/THz) channels, and discusses the solutions proposed in the literature. Section~\ref{sec:wideband-estimation} presents estimation techniques for RIS-aided orthogonal frequency-division multiplexing (OFDM) in wideband channels. Section~\ref{sec:multiuser} presents various approaches for reducing estimation overhead. Section~\ref{sec:machine-learning-methods} presents estimation approaches based on machine learning. Section~\ref{sec:practical-issues} outlines practical issues of contemporary interest and investigation in RIS channel modeling.

\begin{table*}
\caption{Nomenclature}
\label{table:nomenclature}
\centering
\begin{tabular}{|cl|}
\hline
$\ostar$ & circular convolution\\
$\otimes$ & Kronecker product\\
$\odot$ & Hadamard product (element-wise product of two matrices)\\
$\diamond$ & Khatri-Rao product (column-wise Kronecker product)\\
$\mathbf 1$ & vector of all-ones\\
$^T$ & matrix transpose\\
$^*$ & complex conjugate\\
$^H$ & conjugate transpose (Hermitian)\\
$\dagger$ & psuedo-inverse\\
$\diag(\cdot)$ & diagonal matrix built from the argument\\
$\vec(\cdot)$ & concatenation of columns of an $M\times N$ matrix into a vector of size $MN$\\
$||\cdot||$ & vector norm\\
$\mathbb E$ & Expected value\\
$\mathbf \Phi$ & RIS reflection coefficient matrix\\
$\mathbf H$, $\mathbf G$ & incoming \& outgoing RIS channel\\
${\mathbf H}_c$ & end-to-end RIS-induced channel\\
${\mathbf H}_d$ & direct (not through RIS) channel\\
${\mathbf Z}$ & Training matrix, includes pilots \& RIS training states\\
\hline
\end{tabular}
\end{table*}

\section{System and Channel Models}
\label{sec:System-Model}
For easy reference, the mathematical terms and the key variables used in this paper are summarized in Table~\ref{table:nomenclature}.

\subsection{Narrowband/Frequency-Flat Channels}
\label{sec:Narrowband-Model}
Consider a narrowband point-to-point communication between a transmitter with $M_t$ antennas and a receiver with $M_r$ antennas, assisted by an RIS with $N$ elements. Let $\TxRisChannel \in \mathbb{C}^{N\times M_t}$ denote the channel between the transmitter and RIS, $\RisRxChannel \in \mathbb{C}^{M_r \times N}$ the channel between the RIS and the receiver, and $\TxRxChannel \in \mathbb{C}^{M_r \times M_t}$ the direct channel between the transmitter and the receiver. 
The RIS reflection is represented with a matrix $\RisPhaseMatrix = \text{diag}(\psi_1, \psi_2, \dots, \psi_N)$ whose passive elements $\psi_i=\beta_ie^{j\phi_i}$ have amplitude $\beta_i \in [0,1]$.
Then, the received signal is given by
\begin{equation} 
    \mathbf{y}=\sqrt{\rho}(\TxRxChannel+\RisRxChannel\RisPhaseMatrix \TxRisChannel)\inputvec+\noisevec,
    \label{eq:ris-basic-model}
\end{equation}
where the transmit signal $\inputvec\in\mathbb{C}^{M_t \times 1}$ satisfies $\mathbb{E}\|\inputvec\|^2=1$, the transmit power is $\rho$, and the receiver Gaussian noise is $\noisevec \in \mathbb{C}^{M_r\times 1}$.

In the microwave L and S bands, and parts of the C band,\footnote{The frequency boundaries of the rich scattering  model are subject to a number of variables, which are extensively covered in the channel modeling literature~\cite{1033686,1175470,656151}.} a suitable model of rich scattering involves many multipath components, resulting in independent and identically distributed (i.i.d.) Gaussian gains for the direct and RIS-aided channels.
When a direct line-of-sight (LoS) path exists between any two nodes, a Rician fading model applies. In that case, the Tx-RIS channel can be decomposed as follows:
\begin{equation}
    \TxRisChannel = \sqrt{\frac{K_g}{1+K_g}}\bar{\TxRisChannel} + \sqrt{\frac{1}{1+K_g}}\widetilde{\TxRisChannel},
    \label{eq:RicianModel}
\end{equation}
where $K_g$ is the Rician factor, $\bar{\TxRisChannel}$ is the deterministic  LoS component, and $\widetilde{\TxRisChannel}$ is the random non-LoS component. 

Recently, RISs have been studied in higher frequencies (mmWave and THz) where the channels are represented in parametric form; for example, a transmitter-RIS channel with  $L$ resolvable paths is given by 
\begin{equation}
    \TxRisChannel=\sum_{\ell=0}^{L-1}\alpha_\ell \mathbf{a}_2(\phi_{2,\ell},\theta_{2,\ell})\mathbf{a}_1^H(\phi_{1,\ell},\theta_{1,\ell}),
    \label{eq:Sparse-Channel}
\end{equation}
where $\alpha_\ell$ is the complex gain of path $\ell$, $\mathbf{a}_1(\cdot)$ and $\mathbf{a}_2(\cdot)$ are steering vectors at the transmitter and RIS, respectively,
with $\phi_{1,\ell}(\theta_{1,\ell})$ and $\phi_{2,\ell}(\theta_{2,\ell})$ being azimuth (elevation) angles of departure and arrival for the path $\ell$. In matrix form:
\begin{equation}
\TxRisChannel= {\mathbf A}_2 \; \diag({\boldsymbol \alpha}) \; {\mathbf A}_1^H
\label{eq:GeometricModel}
\end{equation}
This matrix representation will be revisited in Section~\ref{sec:sparse-estimation} for channel estimation.

\subsection{Wideband/Frequency-Selective Channels}
\label{sec:Wideband-Model}
Consider an RIS-aided communication between a single antenna transmitter and receiver (for ease of exposition) in frequency selective channels using OFDM modulation with $M_c$ orthogonal sub-carriers. Let $L$ denote the number of channel taps for both the direct and cascaded channels\footnote{Assumed for ease of exposition. Extensions for the case where direct and cascaded channels have a different number of paths is straightforward.}, with $\mathbf{h}_d=[h_d(0),\dots,h_d(L-1),\mathbf{0}_{1\times(M_c-L)}]^T\in\mathbb{C}^{M_c\times1}$ being the zero-padded (time-domain) direct channel vector, $\TxRisChannelvec_n = [g_n(0), \dots, g_n(L-1), \mathbf{0}_{1\times (M_c-L)}]^T$ the channel between transmit antenna and the RIS element $n$, and $\RisRxChannelvec_n = [h_n(0), \dots, h_n(L-1), \mathbf{0}_{1\times (M_c-L)}]^T$ the channel between RIS element $n$ and receive antenna. Let $\widecheck{\inputvec}$ 
be the $M_c \times 1$ transmit frequency-domain OFDM symbol and $\inputvec$ 
be its $M_c$-point inverse discrete Fourier transform (IDFT). At the OFDM transmitter, $\widecheck{\inputvec}$ is first transformed to $\inputvec$ and transmitted over the frequency-selective channel. The signal reaching the receiver due to direct path is given by $\inputvec \ostar \mathbf{h}_d$, where $\ostar$ denotes the circular convolution. The signal reaching the receiver via reflection from RIS element $n$ is  $\psi_n(\inputvec \ostar \TxRisChannelvec_n)\ostar \RisRxChannelvec_n$, where $\psi_n$ is the induced reflection coefficient. The overall received signal due to the direct path and the RIS reflections is 
\begin{align}
    \outputvec = \sqrt{\rho}\left(\sum_{n=1}^N \psi_n(\inputvec \ostar \TxRisChannelvec_n)\ostar \RisRxChannelvec_n + \inputvec \ostar \mathbf{h}_d\right) + \noisevec.
    \label{eq:OFDM-rx-signal-time}
\end{align}
The OFDM receiver performs DFT operation on $\outputvec$ to obtain $\widecheck{\outputvec} =\mathbf{F}_{M_c}\outputvec$, where $\mathbf{F}_{M_c}$ is the $M_c\times M_c$ DFT matrix. Let $\widecheck{\mathbf{h}}_d = \mathbf{F}_{M_c}\mathbf{h}_d$ denote the frequency domain Tx-Rx channel. Similarly, let $\widecheck{\TxRisChannelvec}_n$ and $\widecheck{\RisRxChannelvec}_n$ denote the frequency domain RIS-aided channels due to element $n$. Then, the received signal in the frequency domain is given by 
\begin{align}
    \widecheck{\outputvec} &= \sqrt{\rho} \left( \sum_{n=1}^N \psi_n (\widecheck{\inputvec}\odot  \widecheck{\TxRisChannelvec}_n) \odot \widecheck{\RisRxChannelvec}_n  + \widecheck{\inputvec} \odot \widecheck{\mathbf{h}}_d \right) +  \noisevec \nonumber \\
    &=\sqrt{\rho} \mathbf{X} \left( \sum_{n=1}^N \psi_n (\widecheck{\TxRisChannelvec}_n \odot \widecheck{\RisRxChannelvec}_n) + \widecheck{\mathbf{h}}_d \right) +  \noisevec,
\label{eq:RIS-OFDM-Model1}    
\end{align}
where $\mathbf{X} = \diag(\widecheck{\inputvec})$ and $\odot$ denotes the Hadamard (elementwise) product. The channel gains can follow either rich or sparse scattering models depending on the propagation environment and frequency of operation.

\section{Pilots, Training States, and Linear Estimation}
\label{sec:linear-estimation}
A narrowband RIS-aided system is modeled by Eq.~\eqref{eq:ris-basic-model}, or alternatively, 
\begin{align}
\outputvec = \sqrt{\rho}(\inputvec^T \otimes \identity_{M_r})\vec(\TxRxChannel + \RisRxChannel\RisPhaseMatrix\TxRisChannel) + \noisevec.
\label{eq:alternate-model}
\end{align}
Recall $\otimes$ and $\vec(\cdot)$ denote Kronecker product and vectorization, respectively. Define
\begin{equation}
\CascadedChannel \triangleq [\vec(\TxRxChannel) \;\; \TxRisChannel^T \!\!\diamond \RisRxChannel]
\label{eq:CombinedChannel}
\end{equation}
where $\diamond$ is a columnwise Kronecker product, which is a special case of the Khatri-Rao product. We further define $\CascadedChannelvec \triangleq \vec(\CascadedChannel)$.
We organize the individual reflection coefficients  $\psi$ into a vector $\RisPhasevec$, which carries the same information as the diagonal matrix $\RisPhaseMatrix=\diag(\RisPhasevec)$. Further, define $\tilde{\RisPhasevec} \triangleq \begin{bmatrix} 1 \\ \RisPhasevec \end{bmatrix}$.
Then, Eq.~\eqref{eq:alternate-model} is expressed as follows
\begin{align}
    \outputvec &= \sqrt{\rho}(\inputvec^T \otimes \identity_{M_r})\CascadedChannel \tilde{\RisPhasevec} + \noisevec \nonumber \\
              & = \sqrt{\rho}(\tilde{\RisPhasevec}^T \otimes \inputvec^T \otimes \identity_{M_r})\CascadedChannelvec + \noisevec \nonumber \\
              &\triangleq  \sqrt{\rho} \mathbf{Z} \CascadedChannelvec + \noisevec,
              \label{eq:ris-estimation-model}
\end{align}
where $\mathbf Z$ represents a matrix that includes the transmit vector as well as the RIS training state, $\CascadedChannelvec$ represents the overall channel (including both cascaded and direct channels), whose estimation is necessary for coherent detection at the receiver and beamforming at the transmitter and RIS. The transmission frame includes a training phase and a data transmission phase. During the training phase, $j=1, \ldots, J$, the pilot signals $\inputvec_j$ and RIS training states $\RisPhasevec_j$ give rise to the overall training matrix ${\mathbf Z}_j =\tilde{\RisPhasevec}_j^T \otimes \inputvec_j^T \otimes \identity_{M_r}$, at the receiver resulting in $\outputvec_j  = \sqrt{\rho} \mathbf{Z}_j \CascadedChannelvec + \noisevec$.

\subsection{Linear Estimation}
Since $\outputvec_j$ is $M_r$-dimensional and $\CascadedChannelvec$ is $M_rM_t(N+1)$-dimensional, the least squares and  minimum mean-square error (MMSE) estimation requires at least $J = M_t(N+1)$ pilots. 
Define:
\[
\tilde{\outputvec}\triangleq\begin{bmatrix}
{\outputvec}_1\\
\vdots\\
{\outputvec}_J
\end{bmatrix} \qquad
\widetilde{\mathbf Z}\triangleq\begin{bmatrix}
{\mathbf Z}_1\\
\vdots\\
{\mathbf Z}_J
\end{bmatrix}
\]
Then, the least squares estimate of $\CascadedChannelvec$ is given by 
\begin{align}
    \CascadedChannelvecEst = \frac{1}{\sqrt{\rho}}\widetilde{\mathbf{Z}}^{\dagger}\tilde{\outputvec},
    \label{eq:LS-estimate}
\end{align}
where $\widetilde{\mathbf{Z}}^{\dagger}$ is the pseudo-inverse of $\widetilde{\mathbf{Z}}$.
Let $\mathbf{R}_{h_c}$ denote the covariance matrix of $\CascadedChannelvec$. Then, the LMMSE estimate of $\CascadedChannelvec$ is given by 
\begin{align}
    \CascadedChannelvecEst = \sqrt{\rho}\mathbf{R}_{h_c}\widetilde{\mathbf{Z}}^H(\rho\widetilde{\mathbf{Z}}\mathbf{R}_{h_c}\widetilde{\mathbf{Z}}^H + \identity_{M_rJ})^{-1}\tilde{\outputvec}.
    \label{eq:LMMSE-estimate}
\end{align}

\begin{figure}
    \centering
    \includegraphics[width=2 in]{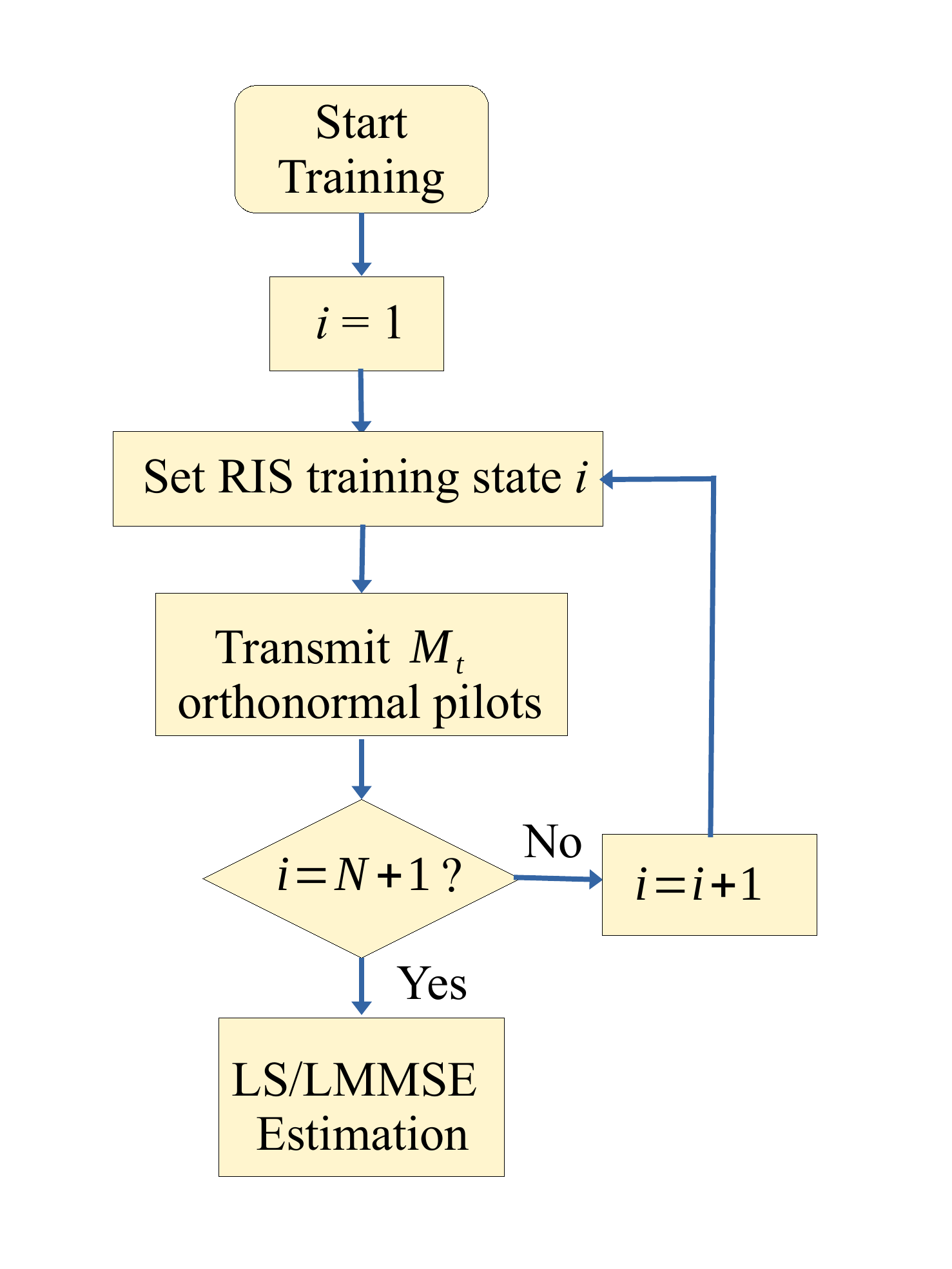}
    \vspace{-0.5 cm}
    \caption{Flowchart representing the channel training process in RIS-aided MIMO.}
    \label{fig:RIS-TrainingFlow}
\end{figure}

{\em Training Process:} As illustrated in Fig. \ref{fig:RIS-TrainingFlow}, the channel training occurs by the RIS assuming a training state $\RisPhasevec_i$ that remains fixed over $M_t$ consecutive slots, and the transmitter emitting $M_t$ linearly independent (preferably orthonormal) pilots. This process repeats $N+1$ times with different RIS training states\footnote{because of $N$ RIS elements and one direct path, constituting $N+1$ degrees of freedom.} that generate linearly independent {\em extended} training vectors $\widetilde\RisPhasevec$. 

\subsection{RIS Training States}
The accuracy of channel estimates depends on the choice of RIS training states. 
We consider the canonical, DFT, and Hadamard training states. For simplicity of exposition, it is assumed that the direct path is absent, and the channel gains follow i.i.d. $\mathcal{CN}(0,1)$ resulting in $\mathbf{R}_{h_c} = \identity_{M_tM_rN}$. The following identity is used in the derivation of least squares and MMSE estimators.
\begin{equation}
\widetilde{\mathbf{Z}}^H\widetilde{\mathbf{Z}} = M_t (\bPsi\bPsi^H)^*\otimes \identity_{M_rM_t},
\label{eq:TrainingMatrixIdentity}\end{equation}
where $\bPsi \triangleq [\RisPhasevec_1 \cdots \RisPhasevec_N]$.
\subsubsection{Canonical Training} 
Canonical training activates one RIS element in each training state, deactivating the remaining $N-1$ elements.\footnote{Nulling the reflection of an RIS element with respect to all angles has not been adequately addressed in the literature and remains an open issue. See Section~\ref{sec:practical-issues} for further details.} Thus, the RIS training vectors constitute a so-called {\em standard basis} or {\em canonical basis}, as follows:
\[
\RisPhasevec_i = {\mathbf e}_i \triangleq [ \delta_{i,1} \; \ldots\;  \delta_{i,N}     ] \qquad i=1,\ldots, N
\]
where $\delta_{i,j}$ is the Kronecker delta function.
This results in $\bPsi\bPsi^H = \identity_N$, and hence, by the identity in Eq.~\eqref{eq:TrainingMatrixIdentity}, $\tilde{\mathbf{Z}}^H\tilde{\mathbf{Z}} = M_t \identity_{M_tM_rN}$. Therefore, from \eqref{eq:LS-estimate}, the least squares estimate with canonical training states is 
\begin{align}
    \CascadedChannelvecEst = \frac{1}{M_t\sqrt{\rho}}\widetilde{\mathbf{Z}}^H \tilde{\outputvec}.
\end{align}
From \eqref{eq:LMMSE-estimate}, the MMSE estimate with canonical training is  
\begin{align}
    \CascadedChannelvecEst = \frac{1}{M_t\sqrt{\rho}\left(1 + \frac{1}{\rho M_t} \right)}\widetilde{\mathbf{Z}}^H \tilde{\outputvec}.
\end{align}

\subsubsection{DFT Training}
In DFT training, each RIS training state is a column of the standard $N\times N$ DFT matrix. The orthogonality $\bPsi\bPsi^H = N \identity_N$ combined with the identity~\eqref{eq:TrainingMatrixIdentity} gives $\widetilde{\mathbf{Z}}^H\widetilde{\mathbf{Z}} = M_t N \identity_{M_tM_rN}$. The least squares estimate with DFT training states is therefore  
\begin{align}
    \CascadedChannelvecEst = \frac{1}{M_t N\sqrt{\rho}}\widetilde{\mathbf{Z}}^H \tilde{\outputvec}.
    \label{eq:LS-DFT-Training}
\end{align}
The MMSE estimate is  
\begin{align}
    \CascadedChannelvecEst = \frac{1}{M_t N\sqrt{\rho}\left(1 + \frac{1}{\rho M_t N} \right)}\widetilde{\mathbf{Z}}^H \tilde{\outputvec}.
\end{align}

\subsubsection{Hadamard Training}
Another choice of RIS training states is to use the columns of $N\times N$ Hadamard matrix, which again results in $\bPsi\bPsi^H = N \identity_N$ and therefore $\widetilde{\mathbf{Z}}^H\widetilde{\mathbf{Z}} = M_t N \identity_{M_tM_rN}$ via the identity~\eqref{eq:TrainingMatrixIdentity}.  The least squares and MMSE estimators with Hadamard training are the same as with DFT training and hence are omitted for brevity.

\begin{remark}
(Ambiguity Problem)\\ 
For any diagonal matrix $\mathbf{D}\in\mathbb{C}^{N\times N}$, $\TxRisChannel'\triangleq\mathbf{D}^{-1}\TxRisChannel$, and $\RisRxChannel'\triangleq\RisRxChannel\mathbf{D}$, 
\begin{equation}
    \RisRxChannel\RisPhaseMatrix\TxRisChannel=\RisRxChannel'\RisPhaseMatrix\TxRisChannel'.
\end{equation}
Therefore, $\TxRisChannel$ and $\RisRxChannel$ cannot be uniquely resolved by observing pilots via the channel $\RisRxChannel\RisPhaseMatrix\TxRisChannel$. 
However, as noted from Eq. \eqref{eq:ris-estimation-model}, the knowledge of cascaded channel $\CascadedChannelvec$ is sufficient for RIS beamforming, while also being more efficient to estimate compared with estimating $\TxRisChannel$ and $\RisRxChannel$ separately. 
\end{remark}

\begin{remark}
(Channel Training under Finite Precision Reflection Coefficients)\\
The precision of RIS coefficients may be limited in practical implementations, with only a finite number of quantized phase shifts being available, affecting both training and beamforming. 
Quantization of phase shifts has no effect on channel estimation under canonical training states.
Hadamard training requires only 1-bit phase precision, without any compromise in estimation accuracy. DFT-training, however, requires $N$ phase shifts, which may not be available in practice for a large RIS array. When $L$ phase shifts ($L < N$) are available at the RIS, a quantized-DFT training can be achieved by mapping the phases of the DFT matrix $\mathbf{F}_N$ to the nearest phases in the quantized phase set $\mathcal{P}$ to obtain quantized-DFT matrix $\mathbf{F}_{q,N}$ as follows:
\begin{align}
\angle{\mathbf{F}_{q,N}(k,\ell)} = \argmin_{\phi \in \mathcal{P}} |e^{-j\phi}  - e^{-j\frac{2\pi(k-1)(\ell-1)}{N}}|.
\end{align}
Unfortunately, the quantized-DFT matrix is non-orthogonal, resulting in degraded channel estimates. This is especially an issue for low-precision RIS implementations. 
\end{remark}

\begin{remark}
(Grouping the RIS Elements)\\
Under some scenarios, the estimation of all channel gain parameters induced by the RIS may be prohibitive in terms of time, power, or both. A remedy has been proposed~\cite{9729398, 9677923, 8937491,9039554, 9467371, 9133184, 9691474, 9292080} that constrains groups of RIS elements to have the same reflection coefficient. Then, it is not difficult to see that the relevant estimation parameter is an aggregate channel gain corresponding to the total reflection produced by the RIS elements in each group (that have the identical reflection coefficient). This scenario is effectively similar to an RIS with fewer (virtual) reflective elements. Each of these virtual elements, representing multiple physical elements, will have a stronger reflection and therefore a stronger channel gain.
An example of this kind is discussed in Section~\ref{sec:wideband-estimation}.
\end{remark}

\section{Estimation vs. Spectral Efficiency}
\label{sec:SE-Training}
Because the RIS channel is not known ahead of time, channel resources must be spent to train and acquire the channel state information. Pilots require transmit power, and transmission time is occupied for generating independent channel observations, commensurate with the number of channel parameters being estimated. Any channel resource used for training becomes unavailable for data transmission. A larger RIS can improve beamforming gain that is beneficial for capacity, but also requires more training resources, which is detrimental for capacity.
This gives rise to an interesting and important tradeoff in the size of RIS and its effects on spectral efficiency, studied in this section.

This section presents the training-based spectral efficiency results for RIS-aided single-antenna transmitters and receivers without a direct path, whose insights carry over to multiple antenna systems as well. The developments in this section follow~\cite{Shamasundar:ISIT22}.
Let $T$ be the coherence interval of all channels, also used as block length.  $T_d$ channel symbols are dedicated to data transmission. In the absence of a direct path, a minimum of $N$ temporal degrees of freedom are needed for training,
\[
T=N + T_d.
\]
Let $\rho_\tau$ and $\rho_d$ denote the training and data powers, respectively, and let $\rho$ denote the average power. Then, by conservation of energy:
\[
\rho T = \rho_\tau N + \rho_d T_d.
\]
With these conditions, the following rate is achievable under canonical training  \cite{Shamasundar:ISIT22}:
\begin{equation}
    R_\tau = \bigg(1 - \frac{N}{T} \bigg)\mathbb{E}_{\CascadedChannelvecEst} \log_2 \bigg( 1 + \frac{\rho_d \big(\sum_{i=1}^N |\CascadedChannelvecEst(i)| \big)^2}{1 + \frac{N\rho_d}{1 + \rho_\tau}}\bigg).
    \label{eq:TrainingBasedCapacityCanonical}
\end{equation}
Under DFT training and Hadamard training, the following rate is achievable:
\begin{equation}
    R_\tau = \bigg(1 - \frac{N}{T} \bigg)\mathbb{E}_{\CascadedChannelvecEst} \log_2 \bigg( 1 + \frac{\rho_d \big(\sum_{i=1}^N |\CascadedChannelvecEst(i)| \big)^2}{1 + \frac{N\rho_d}{1 + N \rho_\tau}}\bigg).
    \label{eq:TrainingBasedCapacityDFT}
\end{equation}
The spectral efficiency is a function of $\rho_\tau$ and $\rho_d$; the optimizer of Eq.~\eqref{eq:TrainingBasedCapacityCanonical} is  $\rho_d T_d = \beta_1^* \rho T$ and $\rho_\tau N = (1-\beta_1^*) \rho T$, with
\begin{equation}
    \beta_1^* = \frac{\sqrt{\big(1 + \frac{\rho T}{N} \big)\big(1 + \frac{N\rho T}{T-N} \big)} - \big(1 + \frac{\rho T}{N} \big)}{\big(\frac{N\rho T}{T-N} - \frac{\rho T}{N}\big)}.
    \label{eq:OptCanonicalTraining}
\end{equation}
Similarly, the optimizer of Eq.~\eqref{eq:TrainingBasedCapacityDFT} is $\rho_d T_d= \beta_2^* \rho T$ and $\rho_\tau N = (1-\beta_2^*) \rho T$, with
\begin{equation}
\beta_2^*=  \frac{\sqrt{(1+\rho T)\big(1 + \frac{N\rho T}{T-N}\big)} - (1 + \rho T)}{\big(\frac{N\rho T}{T-N}  - \rho T\big)}.  
\end{equation}

\begin{figure}
    \centering
    \includegraphics[width=\Figwidth]{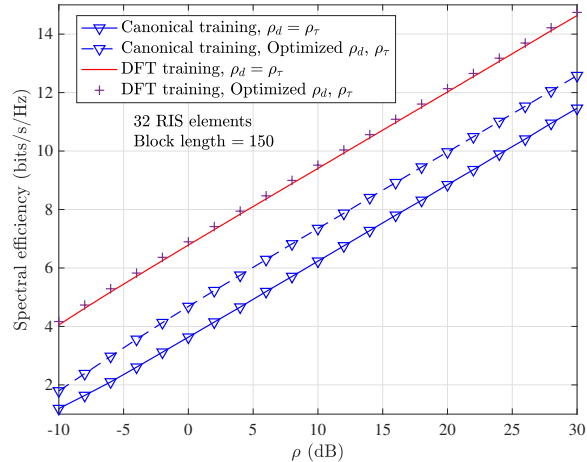}
    \caption{Training-based bounds on capacity with canonical and DFT training.}
    \label{fig:CapacityVsSNR}
\end{figure}
Figure \ref{fig:CapacityVsSNR} shows the training-based bounds on the capacity of RIS-assisted system with 32 RIS elements when the channel coherence interval is $T=150$. The spectral efficiency with DFT training is higher compared with canonical training.\footnote{Hadamard training capacity expressions and numerical results are the same as DFT training, and are omitted for brevity.} Specifically, with equal power allocation between training and data, DFT training achieves a gain of 3.5 bits/s/Hz compared with canonical training. With optimal power allocation for both, DFT training achieves a gain of 2 bits/s/Hz over canonical training. 
The reason for under-performance of canonical training is that the magnitude of RIS training states multiplies the pilot power, therefore the zero coefficients in canonical training reduce the received signal-to-noise ratio for pilots, and induce  a penalty.
On the other hand, DFT training activates all the RIS elements in each training time slot, thereby efficiently utilizing the available pilot power. 

\begin{figure}
    \centering
    \includegraphics[width=\Figwidth]{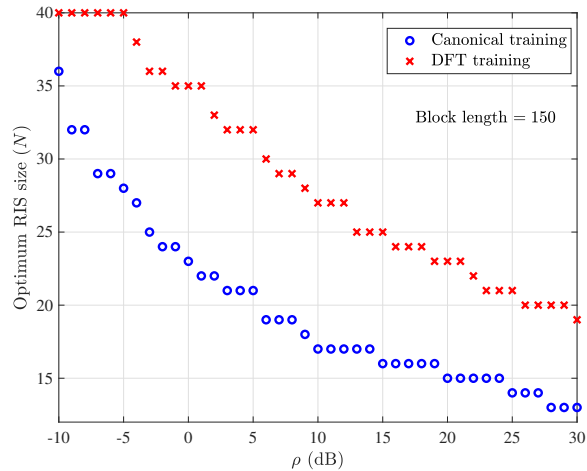}
    \caption{Optimum RIS size that maximizes spectral efficiency.}
    \label{fig:OptimumSize}
\end{figure}
Figure \ref{fig:OptimumSize} shows the RIS array size that maximizes the spectral efficiency, at each signal-to-noise ratio (SNR). At low-SNR, power is at a premium, while degrees of freedom are less important. Therefore, it is beneficial to estimate the channel induced by the entire (available) RIS array, even though the training requires degrees of freedom. Conversely, at high-SNR, degrees of freedom are more important, therefore from a capacity perspective it may be beneficial to utilize only part of an available RIS, so that fewer time slots are utilized for training.

\section{Sparse Channel Estimation}
\label{sec:sparse-estimation}

Whenever the channel has a sparse multipath structure, fewer channel parameters need to be estimated, and the overhead incurred in transmission of pilots and feedback of channel coefficients is reduced.
This is especially relevant for higher frequencies (mmWave/THz) wherein the RIS has the most impact. To capture the efficiencies arising from sparse channel structure, the RIS channel estimation is cast in the form of sparse vector recovery and solved with compressive sensing algorithms that reduce the number of required channel measurements compared with traditional channel estimation~\cite{9103231, 9328485}.

Sparse multipath channels are characterized by a geometric model involving angles of arrival/departure and complex gains of the signal paths. The goal of sparse channel recovery is to estimate the parameters of the angular representation of the channel described in Sec. \ref{sec:Narrowband-Model}, Eq.~\eqref{eq:GeometricModel}, i.e.,
\[
\TxRisChannel= {\mathbf A}_2 \; \diag({\boldsymbol \alpha}) \; {\mathbf A}_1^H
\]
which involves angles of arrival/departure captured in the left and right matrix, and the path strengths captured in the diagonal matrix. Even though $\mathbf G$ might be (highly) rank deficient, it is not (yet) expressed in a suitable format for compressive sampling algorithms. In Eq.~\eqref{eq:GeometricModel}, the basis vectors (columns of ${\mathbf A}_1$ and ${\mathbf A}_2$) can take values over an uncountably infinite set. To recast the problem in a friendly format for compressive sampling, the candidate angles are restricted to a finite set of size $G$, often corresponding to a uniform grid in a prescribed coordinate system. The basis vectors\footnote{also known as steering vectors} corresponding to the discretized angles are collected into {\em dictionary matrices} $\widetilde{\mathbf{A}}_1 \in \mathbb{C}^{M_t\times G}$ and $\widetilde{\mathbf{A}}_2 \in \mathbb{C}^{N \times G}$. With sufficiently good quantization of angles, matrices ${\mathbf A}_1$ and ${\mathbf A}_2$ are approximately\footnote{because the true AoA/AoD may not fall exactly on the quantized grid of angles} submatrices of $\widetilde{\mathbf{A}}_1$ and $\widetilde{\mathbf{A}}_2$. A sparse 
matrix $\bLambda_g$ can select the appropriate columns from $\widetilde{\mathbf{A}}_1$ and $\widetilde{\mathbf{A}}_2$ so that:
\begin{equation}
\TxRisChannel= {\mathbf A}_2 \; \diag({\boldsymbol \alpha}) {\mathbf A}_1^H \approx \widetilde{\mathbf A}_2 \; \bLambda_g \widetilde{\mathbf A}_1^H
\end{equation}
The problem is now in a standard form for compressive sampling. With pre-determined dictionaries $\widetilde{\mathbf{A}}_1$ and $\widetilde{\mathbf{A}}_2$, the objective is to estimate the 
sparse matrix $\bLambda_g$
from a noisy linear observation of $\mathbf G$ (via pilots).

If the transmitter-to-RIS channel $\mathbf G$ has $L$ paths, the discretized grid of angles must have $G \gg L$ elements. Ignoring for now any grid mismatch issues
\[
    \TxRisChannel= {\widetilde{\bA}}_2\bLambda_g {\widetilde{\bA}}_1^H.
\]
Similarly, let $P$ denote the sparsity of RIS-to-receiver channel $\RisRxChannel$. Consider a discretized set of candidate angles with size $H \gg P$, and collect the steering vectors corresponding to these angles into dictionary matrices $\widetilde{\bB}_1 \in \mathbb{C}^{N \times H}$, $\widetilde{\bB}_2 \in \mathbb{C}^{M_r\times H}$. Then, 
\[
\RisRxChannel = \widetilde{\bB}_2 \bLambda_h \widetilde{\bB}_1^H,
\]
where $\bLambda_h$ is a $H \times H$ sparse matrix with $P$ non-zero elements. For simplicity of exposition, we assume no direct path exists between transmitter and receiver. Thus, the received signal in Eq. \eqref{eq:ris-basic-model} takes the form 
\begin{align}
    \outputvec &= \sqrt{\rho} \widetilde{\bB}_2\bLambda_h \widetilde{\bB}_1^H \RisPhaseMatrix \widetilde{\bA}_2 \bLambda_g \widetilde{\bA}_1^H \inputvec + \noisevec \nonumber \\
    &=\sqrt{\rho} \bar{\mathbf{H}}\inputvec + \noisevec \nonumber \\
    & = \sqrt{\rho} (\inputvec^T \otimes \identity_{M_r}) \vec{(\bar{\mathbf{H}})} + \noisevec,
    \label{eq:Vectorization1}
\end{align}
where $\bar{\mathbf{H}} \triangleq \widetilde{\bB}_2\bLambda_h \widetilde{\bB}_1^H \RisPhaseMatrix \widetilde{\bA}_2 \bLambda_g \widetilde{\bA}_1^H $. Using a series of vectorization operations, it can be shown that
\[
\vec{(\bar{\mathbf{H}})} = (\widetilde{\bA}_1^* \otimes \widetilde{\bB}_2)(\bLambda_g^T \otimes \bLambda_h)(\widetilde{\bA}_2^T \diamond \widetilde{\bB}_1^H) \RisPhasevec.
\]
Substituting in Eq. \eqref{eq:Vectorization1}, 
\begin{align}
    \outputvec \hspace{-1mm}&=\hspace{-1mm} \sqrt{\rho}(\inputvec^T \otimes \identity_{M_r})(\widetilde{\bA}_1^* \otimes \widetilde{\bB}_2)(\bLambda_g^T \otimes \bLambda_h)(\widetilde{\bA}_2^T \diamond \widetilde{\bB}_1^H) \RisPhasevec \hspace{-1mm}+ \hspace{-1mm} \noisevec \nonumber \\
    & = \sqrt{\rho} \mathbf{K}(\inputvec, \RisPhasevec) \blambda + \noisevec,
\end{align}
where $\blambda \triangleq \vec{(\bLambda_g^T \otimes \bLambda_h)}$ is the $(GH)^2 \times 1$ sparse vector with $LP$ non-zero elements, and
\[\mathbf{K}(\inputvec, \RisPhasevec) \triangleq ((\widetilde{\bA}_2^T \diamond \widetilde{\bB}_1^H) \RisPhasevec)^T\otimes ((\inputvec^T \otimes \identity_{M_r})(\widetilde{\bA}_1^* \otimes \widetilde{\bB}_2))\]
is the effective measurement matrix which is a function of pilots and RIS training states. During the training phase, the input sequence $(\inputvec_j, \RisPhasevec_j)$ takes $J$ distinct values, and the output sequence $\outputvec_j$ is observed: 
\begin{align}
\begin{bmatrix}
\outputvec_1 \\ \vdots \\ \outputvec_{J}
\end{bmatrix}
 = \sqrt{\rho}
 \begin{bmatrix}
\mathbf{K}(\inputvec_1, \RisPhasevec_1) \\ \vdots \\ \mathbf{K}(\inputvec_{J}, \RisPhasevec_{J})
\end{bmatrix}
\blambda +
\begin{bmatrix}
\noisevec_1 \\ \vdots \\ \noisevec_{J}
\end{bmatrix}.
\label{eq:Sparse-Estimation}
\end{align}
The sparse channel vector $\blambda$ can be reconstructed from the measurements in \eqref{eq:Sparse-Estimation} using standard sparse recovery algorithms such as orthogonal matching pursuit (OMP)~\cite{9103231} and subspace pursuit~\cite{9091552}. Alternating direction method of multipliers (ADMM)~\cite{9475488} and approximate message passing~\cite{9794650} have also been explored for sparse channel estimation in RIS.
Noh {\em et al.}~\cite{9543577} show that, for an RIS-aided single antenna system employing $J$ pilots ($J<N$) for sparse channel estimation, using the $J$ equi-spaced columns of the $N\times N$ DFT matrix as training states produce lower mean squared error compared with canonical training states and the first $J$ columns of the DFT matrix.

{\em Training Overhead and Complexity}: To reconstruct an $LP$-sparse vector of length $(GH)^2$, orthogonal matching pursuit requires $\mathcal{O}(LP\log(GH))$ measurements~\cite{8577023}.
Since each pilot provides $M_r$ measurements, the required number of pilots for sparse RIS channel estimation is $\mathcal{O}\left(\frac{LP}{M_r}\log(GH)\right)$. Subspace pursuit requires even fewer measurements; specifically, it requires $\mathcal{O}(LP\log(\frac{GH}{\sqrt{LP}}))$ measurements~\cite{8577023}, and hence $\mathcal{O}(\frac{LP}{M_r}\log(\frac{GH}{\sqrt{LP}}))$ pilots. In general, $L$ and $P$ are small at high frequencies, and hence the contribution of $LP$ to the training overhead is small. 
Also, due to the logarithmic dependence on $GH$, the induced overhead is low, especially when $M_r$ is large.
A thorough characterization of optimal training with sparse recovery, and the associated training-based capacity (as in Sec. \ref{sec:SE-Training}) is still open. 
The complexity of estimation using orthogonal matching pursuit (and subspace pursuit) is $\mathcal{O}((LPGH)^2\log(GH))$, while linear estimation has a complexity of $\mathcal{O}((M_tM_rN)^3)$. With suitable choice of $G$ and $H$, sparse recovery can reduce the complexity when compared with linear estimation. 

Beyond sparsity, other structural properties can be exploited for further reducing the training overhead. For broadband channel estimation in RIS-aided mmWave massive MIMO systems, Wan {\em et al.}~\cite{9149146}  exploit the common sparsity shared by different sub-carriers and propose a distributed orthogonal matching pursuit to reduce the overhead.  
In the context of an RIS-aided multiuser downlink setting, Wei {\em et al.}~\cite{9328485} show that the angular cascaded channels associated with different users have exactly the same non-zero rows and some common non-zero columns (termed as double structured sparsity). The adaptation of orthogonal matching pursuit to this double-sparse structure is shown to further reduce overhead. Zhou {\em et al.} \cite{9732214} consider uplink channel estimation in RIS-aided mmWave massive MIMO and exploit the fact that in many scenarios the angles of arrival/departure between RIS and base station remain unchanged over multiple coherence blocks. Therefore, the base-station to RIS channel parameters, which are more numerous in massive MIMO, need fewer updates, which can reduce pilot overhead.
Lin {\em et al.} \cite{9760391} decompose the sparse channel recovery into three components: recovery of angle of arrival, angle of departure, and complex gains. 
A semi-passive RIS with a few receiver chains at the RIS was proposed by~\cite{9370097,9127834}. A semi-passive RIS allows for receiving pilots and channel estimation {\em at the RIS}. The transmitter-RIS channel is estimated with the aid of the few RIS on-site measurements, and utilizing compressive sensing.


Matrix factorization and matrix completion~\cite{8879620} can also be used for estimating rank-deficient, sparse RIS channels. The key idea of this approach can be explained as follows. With pilots $\mathbf{X}_\tau=[\inputvec_1 \dots \inputvec_J]$, RIS training states $\bPsi_\tau = [\RisPhasevec_1 \dots \RisPhasevec_J]$, and received pilots $\mathbf{Y}_\tau = [\outputvec_1 \dots \outputvec_J]$, 
the system model in Eq. \eqref{eq:ris-basic-model} becomes~\cite{8879620}
\begin{equation}
    \mathbf{Y}_\tau= \sqrt{\rho} \RisRxChannel(\bPsi_\tau \odot (\TxRisChannel\mathbf{X}_\tau))+\mathbf{N},
    \label{eq:Matrix-Factorization-1}
\end{equation}
where $\mathbf{N} \in \mathbb{C}^{M_r \times J}$ is the additive noise matrix. Equation \eqref{eq:Matrix-Factorization-1} can be equivalently written in the factored form as
\begin{equation}
     \mathbf{Y}_\tau=\sqrt{\rho} \RisRxChannel\mathbf{A}+\mathbf{N},
     \label{eq:Matrix-Factorization-2}
\end{equation}
 where $\mathbf{A} \triangleq \bPsi_\tau \odot (\TxRisChannel\mathbf{X}_\tau)$. With this representation, the estimation is achieved using a two stage process. In the first stage (matrix factorization), the matrices $\RisRxChannel$ and $\mathbf{A}$ are estimated based on $\mathbf{Y}_\tau$. In the second stage (matrix completion), $\TxRisChannel$ is estimated based on the estimate of $\mathbf{A}$. The success of this method requires $\mathbf{A}$ to be sparse and $\RisRxChannel$ to be a low-rank matrix. The sparsity of $\mathbf{A}$ can be satisfied by selecting the RIS training matrix $\bPsi_\tau$ to be sparse, i.e., most of its coefficients set to zero. High frequency (mmWave/THz) channels $\RisRxChannel$  have low rank due to dominance of reflections over scattering.  He and Yuan~\cite{8879620} achieve the matrix factorization step using the bilinear generalized approximate message passing algorithm \cite{6898015}, and the matrix completion step using Riemannian manifold gradient-based algorithm \cite{wei2016guarantees}. Other methods based on similar ideas can be found in~\cite{9133156,9354904,9403420,9556593,he2021semi}. 

The majority of the literature on sparse channel estimation assume that the true angles of arrival/departure lie on a discretized grid (i.e., on the discrete steering angles of the dictionary matrices). In practice, when the angles of arrival/departure do not coincide with the discrete angles in the dictionary, the sampling process leads to many non-zero sample measurements, degrading the sparse recovery algorithm. The sensitivity of sparse recovery to grid mismatch was systematically analyzed in~\cite{5710590}, but this analysis has not been widely adopted in the sparse channel estimation literature.
In an alternative approach, He {\em et al.}~\cite{9398559} propose atomic norm minimization for RIS channel estimation. Atomic norm is a convex function that generalizes the $\ell_1$ norm for sparse recovery and nuclear norm (i.e., sum of singular values) for low-rank matrix completion. Atomic norm minimization works in the continuous domain and avoids discretization, therefore eliminating the grid mismatch problem~\cite{6576276}. Its solution is often via semidefinite programming.

\section{Wideband Channel Estimation}
\label{sec:wideband-estimation}

The RIS-aided OFDM model was discussed in Sec. \ref{sec:Wideband-Model}, where the frequency domain input-output relation was provided by Eq. \eqref{eq:RIS-OFDM-Model1}. Based on this model, the present section provides the main ideas involved in the channel estimation for RIS-aided OFDM. The system model in Eq. \eqref{eq:RIS-OFDM-Model1} can be equivalently written as
\begin{align}
    \widecheck{\outputvec} &= \sqrt{\rho}\mathbf{X}\left(\mathbf{B} \RisPhasevec + \widecheck{\mathbf{h}}_d\right) + \noisevec, 
    \label{eq:RIS-OFDM-Model2} 
\end{align}
where $\mathbf{B}$ is an $M_c \times N$ matrix whose columns are $\widecheck{\TxRisChannelvec}_n \odot \widecheck{\RisRxChannelvec}_n$. The $M_c \times T$ frequency-time frame is divided into two sub-frames: a training sub-frame of size $M_c \times (N+1)$ and data-transmission sub-frame of size $M_c \times (T - N -1)$. In the training sub-frame, the received pilot sequence is given by  
\[
    \widecheck{\outputvec}_j = \sqrt{\rho}\mathbf{X}_j\left(\mathbf{B} \RisPhasevec_j + \widecheck{\mathbf{h}}_d\right) + \noisevec_j, \ j = 1,\dots, N+1.
\]
Defining  
$\bPsi = \big[\RisPhasevec_1 \cdots \RisPhasevec_{N+1}\big]$, 
and using the pilot sequence $\mathbf{X}_j = \identity_N \text{ for } j = 1,\dots, N+1$, the received training sequence $\widecheck{\mathbf{Y}} = [\widecheck{\outputvec}_1, \dots, \widecheck{\outputvec}_{N+1}]$ is given by 
\[
    \widecheck{\mathbf{Y}} = \sqrt{\rho} \begin{bmatrix}\widecheck{\mathbf{h}}_d & \mathbf{B}\end{bmatrix} \begin{bmatrix} {\mathbf 1}^T \\ \bPsi\end{bmatrix}+ \mathbf{W},
\]
where $\mathbf{W} = [\noisevec_1, \dots, \noisevec_{N+1}]$. 
For convenience of notation we define:
\[
{\mathbf C} \triangleq \begin{bmatrix}\widecheck{\mathbf{h}}_d & \mathbf{B}\end{bmatrix} \qquad \widetilde{\bPsi} \triangleq \begin{bmatrix} {\mathbf 1}^T \\ \bPsi\end{bmatrix} 
\]
resulting in
\begin{equation}
 \widecheck{\mathbf{Y}} = \sqrt{\rho} \; {\mathbf C} \widetilde{\bPsi}+ \mathbf{W}.
\end{equation}
From this, the matrix $\mathbf C$ containing the direct and cascaded channels can be estimated as 
\begin{align}
   \widehat{\mathbf C} = \frac{1}{\sqrt{\rho}}\widecheck{\mathbf{Y}}\widetilde{\bPsi}^{-1}.
\end{align}
It has been shown in \cite{8937491} that choosing $\widetilde{\bPsi} = \mathbf{F}_{N+1}$ results in the least error variance. 

\begin{figure}[h]
    \centering
    \includegraphics[width=3.3 in]{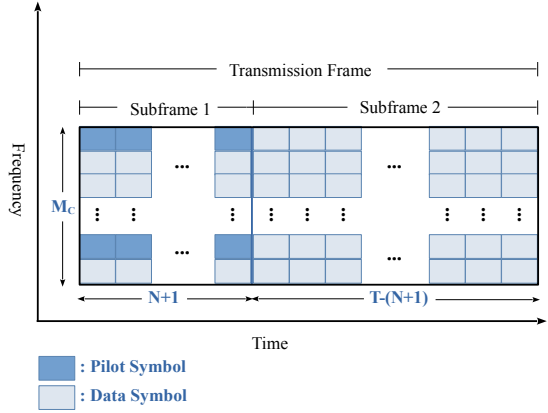}
    \caption{Transmission frame structure in RIS-aided OFDM~\cite{8937491}.}
    \label{fig:Wideband-estimation}
\end{figure}

In the above method, pilots were inserted on all the $M_c$ sub-carriers of the $N+1$ training OFDM symbols. In practice, when the channel is correlated in the frequency domain, fewer pilots may be employed and then channel estimates may be interpolation among sub-carriers.\footnote{This is a long-standing practice that has been adopted in 3GPP.} The system model in Eq. \eqref{eq:RIS-OFDM-Model2} is equivalent to:
\begin{align}
    \widecheck{\outputvec} = \sqrt{\rho}\mathbf{X}\mathbf{q} + \noisevec,    
    \label{eq:RIS-OFDM-Model3}              
\end{align}
where $\mathbf{q} \triangleq \mathbf{B} \RisPhasevec + \widecheck{\mathbf{h}}_d$. Now, as shown in Fig. \ref{fig:Wideband-estimation}, in the training sub-frame, $N_p$ pilots ($N_p < M_c$) are inserted in each OFDM symbol with a spacing of $\Delta = \lfloor \frac{M_c}{N_p} \rfloor$. Let $\mathcal{P} = \{0, \Delta, \dots, (N_p - 1)\Delta \}$ denote the indices of the sub-carries containing pilots. Let $\widecheck{\inputvec}_{\mathcal{P}}$ and $\widecheck{\outputvec}_{\mathcal{P}}$ denote the transmitted and received pilots on sub-carriers indexed by $\mathcal{P}$, respectively. Also, let $\mathbf{X}_{\mathcal{P}} = \diag(\widecheck{\inputvec}_{\mathcal{P}})$. Then, the estimate of $\mathbf{q}_{\mathcal{P}}$ can be obtained as
\[
    \hat{\mathbf{q}}_{\mathcal{P}} = \frac{1}{\sqrt{\rho}}\mathbf{X}_{\mathcal{P}}^{-1} \widecheck{\outputvec}_{\mathcal{P}}.
\]
Using $\hat{\mathbf{q}}_{\mathcal{P}}$, the estimate of $\mathbf{q}$ (denoted by $\hat{\mathbf{q}}$) is obtained via interpolation along the subcarriers. The work in~\cite{8937491} applies the DFT/IDFT-based interpolation on the pilot sequence. 
Now, in order to resolve $\mathbf{B}$ and $\widecheck{\mathbf{h}}_d$ from $\hat{\mathbf{q}}$, the RIS training states $\RisPhasevec_j$ are adjusted during each training OFDM symbol and the corresponding $\hat{\mathbf{q}}_j$ is given by
\[
    \hat{\mathbf{q}}_j = \mathbf{B}\RisPhasevec_j + \widecheck{\mathbf{h}}_d + \mathbf{v}_j,  \ j = 1, \dots, N+1
\]
where $\mathbf{v}_j$ is the error in estimating $\mathbf{q}$ during the pilot slot $j$. Let  $\widehat{\mathbf{Q}} = [\hat{\mathbf{q}}_1 \dots \hat{\mathbf{q}}_{N+1}]$ and $\mathbf{V} = [\mathbf{v}_1, \dots , \mathbf{v}_{N+1}]$. Then, 
\[
    \widehat{\mathbf{Q}} = \mathbf{C} \widetilde{\bPsi} + \mathbf{V}.
\]
Using the above equation, the estimate of matrix $\mathbf{C}$ containing direct and RIS-aided channels can be obtained as 
\begin{align}
    \widehat{\mathbf{C}} = \widehat{\mathbf{Q}}\widetilde{\bPsi}^{-1}. 
\end{align}

To further reduce the estimation overhead, one may group the neighboring RIS elements and use the same reflection coefficient for all the elements in each group. This solution has been suggested for both flat and frequency selective channels~\cite{9729398, 9677923,8937491,9039554, 9467371, 9133184, 9691474, 9292080}.  If the $N$ RIS elements are divided into $N_g$ groups ($N_g < N$) with each group containing $N/N_g$ elements, then the channel estimation requires estimating only $N_g$ aggregated RIS-aided channels corresponding to each group, instead of estimating all the $N$ cascaded channels corresponding to each RIS element. Therefore, the training overhead is reduced from $N+1$ OFDM symbol durations to $N_g+1$ OFDM symbol durations, where each OFDM symbol duration is composed of $(M_c + L_{cp})$ time-slots with $L_{cp}$ being the length of the cyclic prefix. The grouping strategy trades-off accuracy for overhead in order to improve the overall spectral efficiency, which is analytically characterized in~\cite{9729398}. Zheng {\em et al.}~\cite{9195133} extend this estimation technique to multiuser orthogonal frequency-division multiple access (OFDMA) systems. The same authors~\cite{9261597} propose a fast channel estimation scheme for reducing the training-overhead in RIS-aided OFDM. The key idea is to use short OFDM symbols of $M_c'$ sub-carriers ($L \leq M_c' \ll M_c$) during the training phase, which consumes $(M_c' + L_{cp})$ time-slots per OFDM training symbol. This reduces the $(M_c + L_{cp})$ pilots that were required by \cite{8937491}. 

The above works assume an ideal reflection model in which the RIS elements achieve the same amplitude and phase response across the entire OFDM band. Wenhao {\em et al.}~\cite{9064547} show that the practical response of RIS is tightly related to the frequency of the signal. Based on this, Yang {\em et al.}~\cite{9419982} studies channel estimation for RIS-aided OFDM under a practical reflection model and finite precision coefficients. While differing in its modeling, the estimation technique of~\cite{9419982} is similar to  \cite{8937491}, as outlined above, and is omitted for brevity.

\section{Reducing the RIS Estimation Overhead}
\label{sec:multiuser}

We begin by exploring savings in pilots and estimation overhead that arise from the multi-user nature of the RIS channel. An RIS-aided uplink system with $K$ single-antenna users and $M$-antenna base station (BS) must estimate $KMN + KM$ links for RIS beamforming and equalization, which can be prohibitive either under massive MIMO, or in large cells. Linear estimation techniques (Sec.~\ref{sec:linear-estimation}) require $K(N+1)$ pilot transmission slots, growing linearly with the size of RIS. We discuss avenues for reducing the pilot overhead.

\subsection{Common RIS-BS Channel} 
\label{sec:CommonRisBsChannel}
Consider a scenario where a base station is aided by a single RIS for communication with multiple users.
To describe channel estimation in this multi-user scenario, we adapt the system model from Eq.~\eqref{eq:ris-basic-model} for the multi-user uplink channel:
\begin{equation} 
    \mathbf{y}=\sum_{k=1}^K \sqrt{\rho}({\mathbf h}_{dk}+\RisRxChannel\RisPhaseMatrix {\mathbf g}_k)x_k+\noisevec_k,
\end{equation}
where ${\mathbf h}_{dk}$ is the direct channel from a single-antenna user $k$ to a multi-antenna base station, and ${\mathbf g}_k \in \mathbb{C}^{N}$ is the channel from the user $k$ to RIS. Recall that the combined (direct and cascaded) RIS-aided channel gains to be estimated were collected into a single matrix in Eq.~\eqref{eq:CombinedChannel}; a specialization of that matrix for the case of a single-antenna user  is given by:
\begin{equation}
{\CascadedChannel}_k = [{\mathbf h}_{dk} \;\; \RisRxChannel \; \diag({\mathbf g}_k)]
\end{equation}
Among the quantities participating in this expression, the two vectors ${\mathbf h}_{dk}$ and ${\mathbf g}_k$ are distinct for different users, but the BS-RIS matrix $\RisRxChannel$ is common between users. The user-by-user uplink channel estimation requires one pilot for estimating the direct channel and $N$ pilots for estimating the cascaded channel, for a total of $K(N+1)$ total pilot transmission slots. But the commonality of $\RisRxChannel$ among users hints at possible savings in the total number of needed pilot slots, which we now explore.

To begin with, the direct channels for all users is estimated, by deactivating the RIS. This requires one pilot slot per user, but this step may not be crucial, because it is often the absence of a direct path that makes the RIS an attractive choice. In the next step, the cascaded channel $(\RisRxChannel \; \diag({\mathbf g}_1))$ is measured by emitting $N$ pilots from User~1 to the base station. This is accomplished via $N$ successive training states at the RIS, whose details are omitted for brevity. For User~2, we now need to measure $(\RisRxChannel \; \diag({\mathbf g}_2))$.  This new matrix has columns that are co-directional with columns of $(\RisRxChannel \; \diag({\mathbf g}_1))$, thus only the magnitude of each column needs to be measured. For $N$ columns, this requires $N$ new observation samples, however, reception at the multi-antenna base station provides $M$ independent observations per pilot transmission. Therefore, after obtaining $(\RisRxChannel \; \diag({\mathbf g}_1))$, only $\frac{N}{M}$ pilot transmissions are needed per additional user, as long as training states are designed properly. The design of training states that ensure the requisite linearly independent observations has been explained in~\cite{9130088, 9120452}.
When $M>N$, following the above argument, it is easy to see that one pilot per user is sufficient for estimating the cascaded channels of Users $2,\dots,K$. Therefore, the total training overhead of this scheme is given by 
\begin{align}
J = K+N+\max\Big(K-1, \Big\lceil \frac{(K-1)N}{M}\Big\rceil\Big),
\label{eq:OverheadMultiuser}
\end{align}
where $\lceil \cdot \rceil$ denotes the smallest integer bigger than or equal to the argument.
For massive MIMO systems with $M > N$, the overhead $J = 2K + N -1$, meaning that each user beyond the first one requires two pilot slots. This provides significant savings over the $N+1$ slots needed conventionally. Guan {\em et al.}~\cite{9603291} propose a slight modification of this technique, in which a few stationary nodes called the {\em anchor nodes} are assumed to exist in the network. The anchor nodes transmit pilot signals and the base station estimates anchor-RIS-BS channels. Due to the common RIS-BS channel, the User-RIS-BS channels are subsequently estimated with fewer pilot transmissions. Since the anchor nodes are stationary, the estimation of anchor-RIS-BS channels are done less frequently compared with the earlier, single-reference user in~\cite{9130088, 9120452} which was not assumed to be stationary, thus resulting in additional savings. Guo and Lao~\cite{9839429} also explore the possibility of exploiting the common RIS-BS channel without requiring a reference user.

The methods in \cite{9130088, 9120452} estimate the direct channel, and subsequently subtract it from the measurement intended for the cascaded channel, which is not MMSE optimal. A joint estimate of $[{\mathbf h}_{d1} \;\; \RisRxChannel \; \diag({\mathbf g}_1)]$, i.e., User 1's direct and cascaded channels, has a lower mean squared error (MSE).
Wei {\em et al.}~\cite{9559873} propose this joint estimation, and then the remaining user channels are estimated via the same technique as \cite{9130088, 9120452}. 

This modification acknowledges and addresses the propagation of the error in the estimation of ${\mathbf h}_{d1}$ when estimating the cascaded channel of User~1. The estimate of  $\RisRxChannel$ is used for constructing the cascaded channels of other users too, therefore in a sense, the errors committed in estimating the channel of User~1 can propagate into the estimation of other users' channels. However, since User~2 and subsequent users employ fewer pilots than User~1, it is not obvious that their channel measurements can be used to improve the estimate of $\RisRxChannel$.

\subsection{Slowly Varying BS-RIS Channel}
Since the BS and RIS are static, the channel between them varies slowly. In comparison, the BS-user and RIS-user channels are more dynamic because of the mobility of the user. 
The high dimensional, but slowly varying BS-RIS channel can therefore be estimated less frequently, while the low-dimensional BS-user and RIS-user channels are estimated more frequently. To isolate the estimation of BS-RIS link, \cite{9400843} assumes a full-duplex base station. The base station will emit pilots and listen for the reflection from the RIS. The self-interference of the full-duplex reception must be dealt with, and the BS-RIS channel recovered. Given the BS-RIS channel estimate, the direct BS-user channel and the RIS-user channel are estimated conventionally. The latter estimates are more frequent, but also require smaller overhead. If $T_L$ denotes the coherence time of the BS-RIS channel and $T_S$  denotes the coherence time of the BS-user and RIS-user channels such that $T_L = \alpha T_S$, then the overhead of the two-timescale method is
\[
    J = \frac{2(N+1)}{\alpha} + K\Big\lceil \frac{N}{M} \Big\rceil + K.
\]
In practice, $\alpha \gg 1$ and hence the first term is small. For massive MIMO systems with many base station antennas $M > N$, the overhead becomes $\frac{2(N+1)}{\alpha} + 2K$, i.e., after estimating the BS-RIS channel, each user needs two pilots. Under $\alpha > 2$, this method has smaller overhead compared with the method of Section~\ref{sec:CommonRisBsChannel}, although one must be careful that the two methods address different channels and different base station capabilities, so they are not directly comparable.

\subsection{Infrequent RIS Coefficient Updates}
\label{sec:statistical-csi}
Another source of potential savings in RIS induced channels is to deliberately reduce the frequency with which RIS reflection coefficients are updated. As long as RIS coefficients are not updated, the RIS blends into the channel and effectively the system is reduced to a (multi-user) MIMO system, with conventional channel training and pilots. Of course, this involves a tradeoff: fewer pilot slots are needed, but also, the match of RIS coefficients to the channel will go stale, therefore part of the beamforming gains of RIS will be lost.

Ideally, an analysis of this situation requires a temporally varying channel model with a corresponding temporal correlation. However, the work in this area has taken a different direction, via considering a channel model, with line-of-sight and rich scattering components. For the Tx-RIS channel $\TxRisChannel$, this means the Rician model
\[
   \TxRisChannel = \sqrt{\frac{K_g}{1+K_g}}\bar{\TxRisChannel} + \sqrt{\frac{1}{1+K_g}}\widetilde{\TxRisChannel},
\]
which we also saw in Eq.~\eqref{eq:RicianModel}. A similar model is utilized for the RIS-Rx channels.

It is assumed that the infrequent update of RIS is able to fully capture the line-of-sight component, while not capturing anything about the rich scattering part of the model, even immediately following the pilot transmission. This approximation is different from the common modeling of temporal variance in most wireless channels, in which channel knowledge is accurate at times that are proximate to the pilots, but it has the advantage of removing the complexities involved in the temporal dynamics of the channel. Thus, it reduces the problem to an equivalent problem involving a channel state that is partially known. The literature~\cite{8746155, hu2020statistical, 9198125,9743440, 9500188} refers to this new formulation of channel temporal dynamics as {\em statistical channel state information}.\footnote{This IRS channel model has weak connections with earlier, well-known work in MIMO channels with statistical CSI~\cite{1377920, 1427689} that were driven by CSI impairments due to limited feedback or feedback delay.}

The central idea of~\cite{8746155, hu2020statistical, 9198125,9743440, 9500188} is that the line-of-sight component has a longer coherence interval than the rich scattering component. Single-antenna mobiles estimate the end-to-end uplink channel with a single pilot at the smaller coherence interval, and the base-station beamforming is {\em also} updated at the smaller coherence interval. However, the RIS coefficient is updated only at the longer line-of-sight coherence intervals. This creates significant savings, since most of the pilot slots in the RIS-induced channel, especially for single-antenna mobiles, is needed for estimation and updating of the RIS coefficients.

Several works have attempted to maximize the ergodic downlink rates to single-antenna mobiles with beamforming $\TxBeamformer$, either in the single-user or multiple user scenarios. For the single-user case, the received signal is given by
\[
    y = \sqrt{\rho}(\RisRxChannelvec^H\RisPhaseMatrix\TxRisChannel + \TxRxChannelvec^H)\TxBeamformer s + w,
\]
 The best beamformer $\TxBeamformer$ is found for a given set of RIS coefficients $\RisPhaseMatrix$, but then utilize as $\RisPhaseMatrix$ the (fixed) RIS coefficients that are statistically the best over the variations of the channel.  
\begin{align}
  C^* \hspace{-1mm}=\hspace{-1mm}  \max_{\RisPhaseMatrix}\hspace{0.75mm} \mathbb{E}_{\RisRxChannelvec,\TxRisChannel,\TxRxChannelvec} \hspace{-1mm}\left[\hspace{-0.7mm} \max_{\TxBeamformer} \left\{ \log_2 \left(1 + \rho|(\RisRxChannelvec^H\RisPhaseMatrix\TxRisChannel + \TxRxChannelvec^H)\TxBeamformer|^2 \right) \right\} \hspace{-0.7mm}\right]
  \label{eq:StatisticalCSI-Optimization}
\end{align}
{\color{black} Han {\em et al.}~\cite{8746155} achieve the inner maximization in Eq.~\eqref{eq:StatisticalCSI-Optimization} via maximal ratio transmission, and adjust the reflection coefficients based on an outer bound on the ergodic rate.}   
Hu {\em et al.}~\cite{hu2020statistical} maximize Eq. \eqref{eq:StatisticalCSI-Optimization} via alternating optimization method, Zhao {\em et al.}~\cite{9198125} uses a penalty dual decomposition method, Zhi~{\em et al.}~\cite{9743440} achieve minimum user rate maximization via genetic algorithm, and Gan {\em et al.}~\cite{9500188} propose methods based on ADMM, fractional programming, and alternating optimization.

Several important points and open problems remain for consideration in this area. To begin with, these methods are based on the assumption that the RIS will be changed infrequently, but also calculate and optimize ergodic capacity. Therefore, the practical implementation of these techniques requires an outer code that goes across many coherence intervals of the slower channel. 
In many such cases, outage capacity or throughput may be a more suitable metric for optimization, and there is room for future work in this area. 

Another useful direction is to find simplifications and approximations of the expression in Eq.~\eqref{eq:StatisticalCSI-Optimization} in order to recognize trends and/or suggest different approaches. In this area, there is a need for achievable rate (inner bound) expressions rather than outer bounds. Inner bounds for this expression have not been developed at the time of the writing of this paper. 

\subsection{Opportunistic RIS}
\label{sec:OpportunisticRIS}
Another strategy for reducing the estimation overhead is inspired by an idea that harks back to the concept of opportunistic transmission~\cite{957318,1405296,4290013}. $Q$ randomly selected vectors  $\{ \RisPhasevec_1, \dots, \RisPhasevec_Q\}$ are assigned one-by-one as RIS phase vectors. In each instance, the RIS changes the scattering environment randomly, so there is no beamforming in the usual sense of the word. For each of these $Q$ scattering conditions, the end-to-end multiple-input single-output (MISO) channel is measured using a few pilots, and the best one is chosen for one block of transmission. An and Gan~\cite{9438669} propose the above approach in narrowband channels, and study bounds on its ergodic performance.   The set $\{ \RisPhasevec_1, \dots, \RisPhasevec_Q\}$ is called a codebook in~\cite{9438669}, however, this is a slight misnomer since this set need not be determined or agreed upon ahead of time, is statistically independent of signals emitted from transmit antennas, and is not needed at the receiver for decoding. The connection of this class of techniques with opportunistic transmission is evidenced by the appearance of the order statistics of (induced) channels in~\cite[Proposition 1]{9438669}. An {\em et al.} \cite{9691275} extend this idea to OFDM transmission.

\section{Machine Learning based Channel Estimation}
\label{sec:machine-learning-methods}

Machine learning is being actively investigated for channel estimation; this section explores machine learning channel estimation in the context of RIS.

\begin{figure}
    \centering
    \includegraphics[width=3.35in]{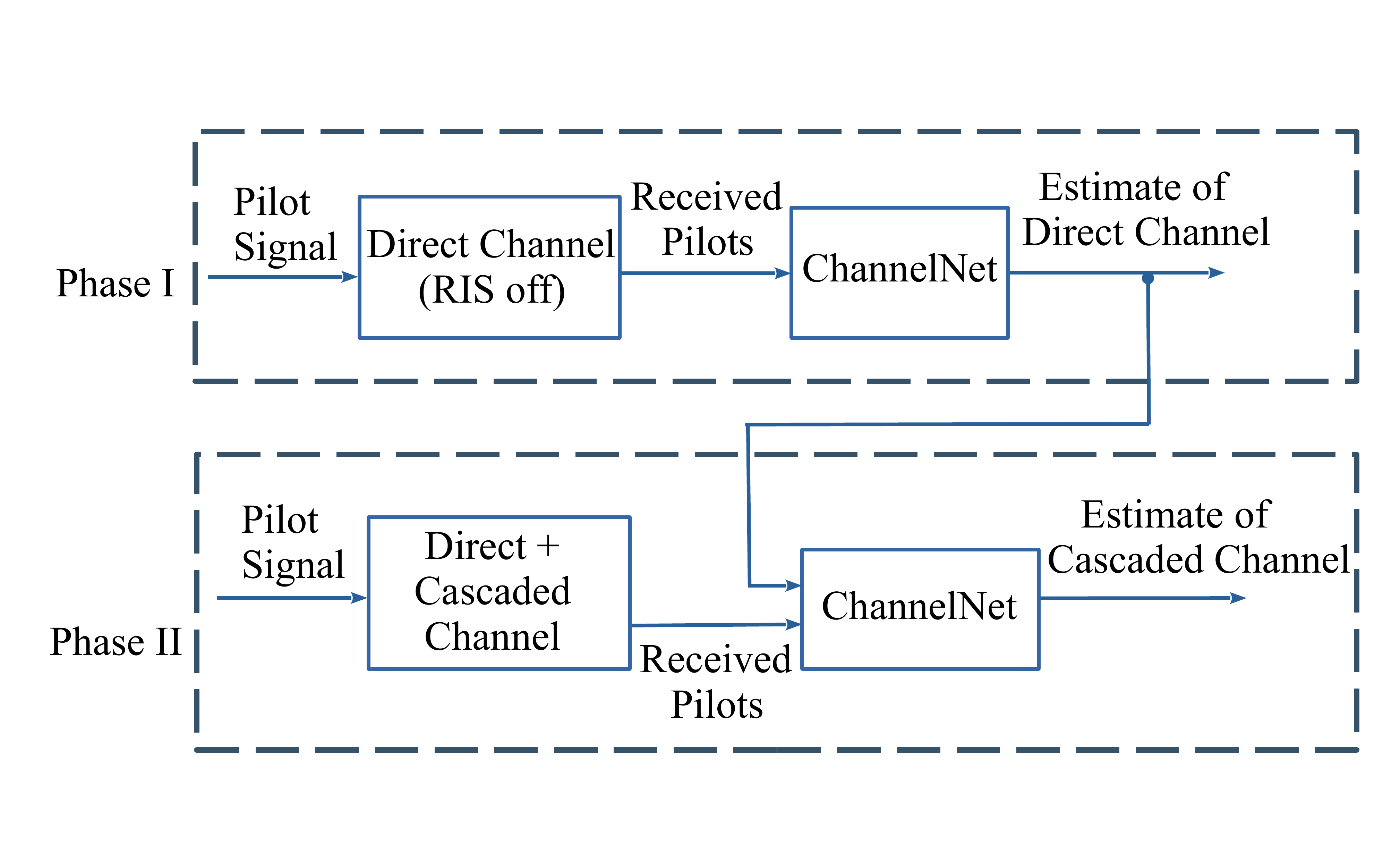}
    \vspace{-1 cm}
    \caption{RIS channel estimation using ChannelNet~\cite{9090876}.}
    \label{fig:ChannelNet}
    \vspace{-0.5 cm}
\end{figure}

Among the early attempts at using machine learning for RIS channel estimation was a non-parametric convolutional neural network estimator by Elbir {\em et al.}~\cite{9090876}, applied to RIS-aided downlink mmWave channels. The estimation is achieved in two phases (see Fig. \ref{fig:ChannelNet}), somewhat similar to other methods seen earlier. In the first phase, the base station transmits pilots while the RIS elements are inactive (turned off) so that the direct channel(s) are estimated at the receivers, each of them operating an instance of the convolutional neural network. In the second phase, the RIS assumes (several) training states while the base station transmits pilots. Each of the users employs the received pilots in this phase, and in combination with the estimated direct channels (obtained in the previous phase), produces an estimate of the cascaded channel using the 
same convolutional neural network architecture. 
Under low SNR conditions, this technique claims better performance than a (corresponding) two-phase least squares technique. Under high SNR conditions, the neural network technique has a performance ceiling, while the least squares techniques do not.  The experiments involved a 64-antenna base station, 100-element RIS, 8 single-antenna users, and a geometric channel with 10 paths. The neural network has an input layer, an output regression layer providing complex valued channel estimates, three convolutional layers each with 256 $3\times 3$ filters, two fully connected layers with 1024 and 2048 nodes. 
The input layer has size $\sqrt{M}\times \sqrt{M}\times 3$  for direct channel estimation and $N\times M\times 3$ for cascaded channel estimation. The output layer has size $2M\times 1$ for direct and $2NM\times 1$ for cascaded channel estimation.

\subsection{Post Processing Least Squares Estimates}
\label{sec:postprocessingLS}
\begin{figure}
    \centering
    \includegraphics[width=3.35 in]{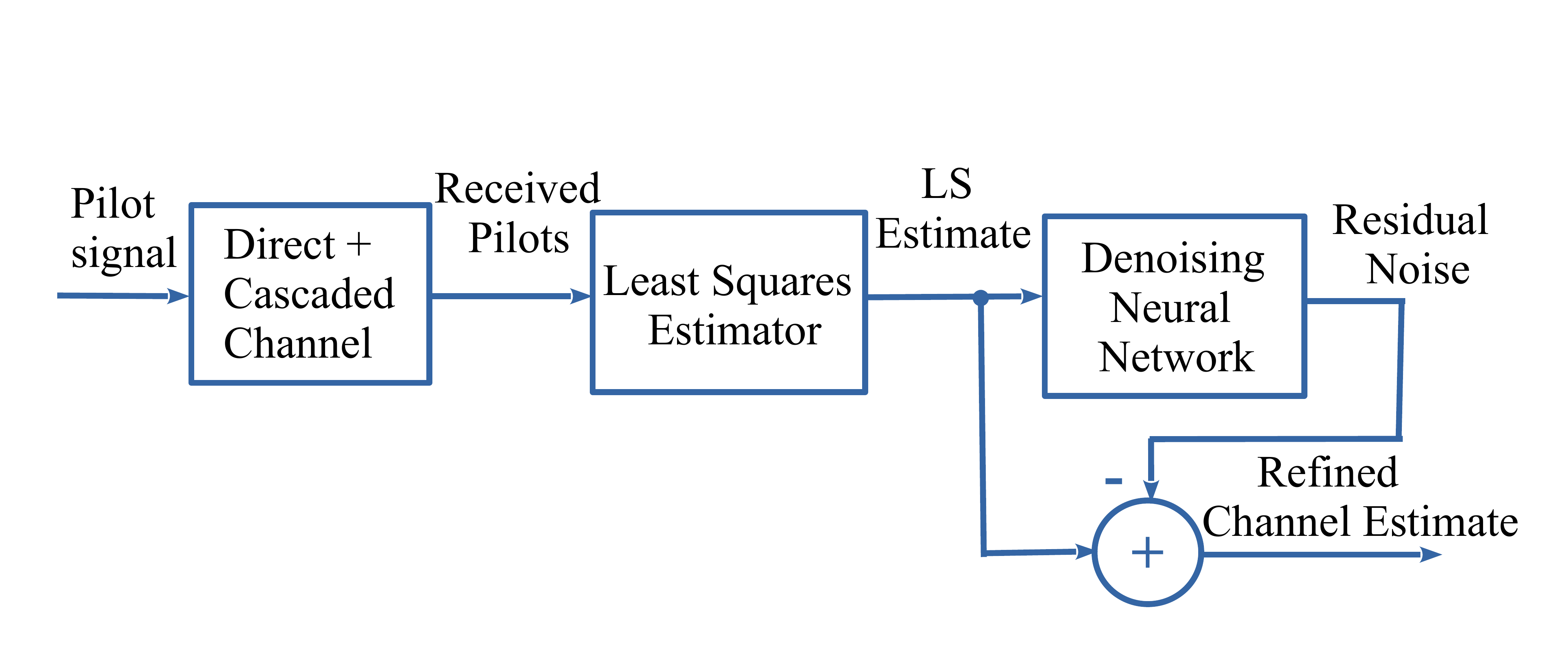}
    \vspace{-0.5 cm}
    \caption{RIS channel estimation via denoising of least squares estimate using neural networks.}
    \label{fig:DenoisingLS}
\end{figure}
Motivated by image denoising using neural networks, Kundu and McKay~\cite{9366894} model the problem of RIS channel estimation as that of denoising the least squares solution (see Fig. \ref{fig:DenoisingLS}). Specifically, in the first step, least squares estimate of direct and cascaded channels is obtained using pilots and DFT training states. The obtained least squares estimate is viewed as a noisy version of the original channel. This is followed by a post-processing step in which the Denoising Convolutional Neural Network (DnCNN)~\cite{7839189} or Fast \& Flexible Denoising Network (FFDNet)~\cite{8365806} are used.\footnote{These neural networks are imported from the image denoising literature.} The least squares channel estimate is the input to the neural network, whose output is an estimate of the least squares estimation error. The post-processed estimate is obtained by subtracting the estimate of the least squares estimation error from the least squares estimate.

The optimal minimum mean squared error estimator of the channel gains is the conditional mean, but since the cascaded channel is non-Gaussian, this estimator is non-linear and difficult to characterize. This has been the main motivation mentioned in~\cite{9366894} for a neural network approach. One can infer that the LMMSE estimate, being linear, is akin to a first order term of a Taylor series expansion for the conditional mean estimator, and the neural network attempts to approximate the higher order terms.

Chang {\em et al.}~\cite{9505267} propose convolutional deep residual networks for denoising the least squares solution. Nipuni {\em et al.}~\cite{9569694} employ neural network denoising for wideband channel estimation in RIS-aided OFDM. Shicong {\em et al.}~\cite{9127834} propose an RIS architecture with a few active elements for initially estimating the low-dimensional channel, compressive sensing reconstruction of the complete high-dimensional channel, and a further refinement with a convolutional neural network. Mao {\em et al.}~\cite{9695448} refine the channel estimates produced by orthogonal matching pursuit using deep residual networks in RIS-aided mmWave channels. Ye {\em et al.} ~\cite{9761227} and Jin {\em et al.}~\cite{9633175} explore generative adversarial networks for estimation in RIS-aided mmWave massive MIMO. 

\subsection{Partial CSI}

\begin{figure*}
    \centering
    \includegraphics[width=6.5 in]{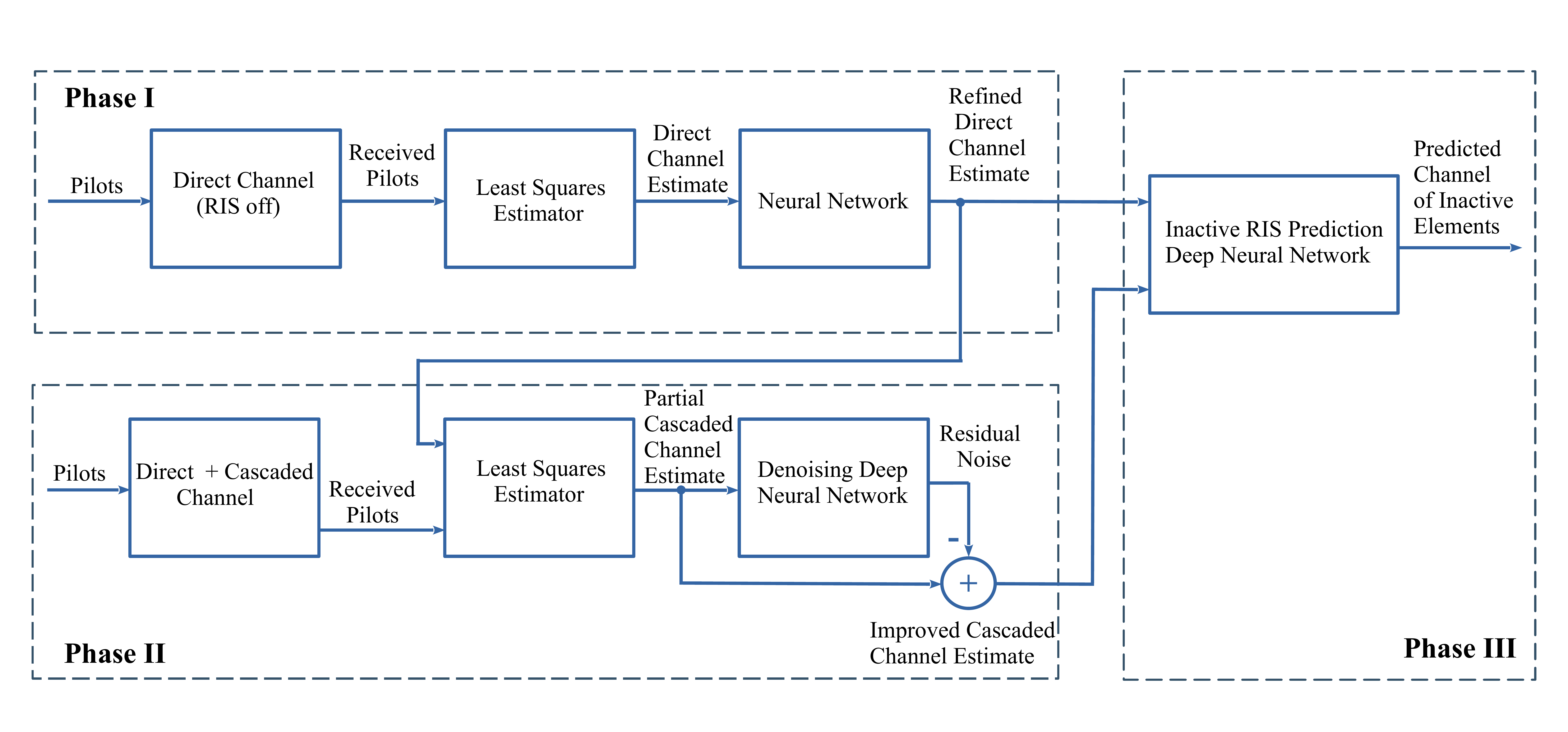}
    \vspace{-0.5 cm}
    \caption{RIS channel estimation using predictive neural networks.}
    \vspace{-0.25 cm}
    \label{fig:ExtrapolatingCSI}
\end{figure*}

In order to reduce the training overhead in deep learning RIS channel estimation, Gao {\em et al.}~\cite{9367208} propose a predictive neural network in the context of RIS-aided uplink massive MIMO. As shown in Fig.~\ref{fig:ExtrapolatingCSI}, the proposed method works in three stages: In the first stage, the RIS is turned off and the direct channel is estimated using least squares method, which is further refined using a fully connected neural network. In the second stage, only a part of RIS elements are activated ($N_1 < N$), and the cascaded channels corresponding to the activated RIS elements are estimated using least squares method with $N_1\times N_1$ DFT training states and further refined using a denoising convolutional neural network. In the third stage, the cascaded channels corresponding to the inactive RIS elements are predicted using a fully connected inactive RIS channel prediction neural network. A geometric channel model is employed for the base-station to RIS channel where the channel matrix is generated from a geometric model that assigns the same gains and angles of arrival/departure to different RIS elements, with an implicit underlying assumption that the scatterers are far from the RIS and the base station.
The proposed estimation needs $N_1 +  1$ pilots instead of $N+1$ pilots, reducing the overhead. 

Xu {\em et al.}~\cite{9373446} propose an ordinary differential equation (ODE) based convolutional neural network for predicting the channel corresponding to inactive RIS elements. 
Shtaiwi {\em et al.}~\cite{9419974} assume a multi-user uplink channel in which the channel of different users is highly correlated, so that only a few users need to transmit pilots, and the channel of the remaining users may be predicted from the first few. A neural network is employed for the prediction.
In RIS-aided uplink communication, Xu {\em et al.}~\cite{9611281} uses spatial correlation between the channels of different RIS elements, as well as temporal correlation of time-varying channels, to reduce the communication overhead. For exploiting spatial correlation, some RIS elements are turned off, hence corresponding channels are not directly estimated by pilots, but are interpolated from other RIS channel estimates using a convolutional neural network. For temporal correlation, a recurrent neural network is used to interpolate the channel values between two pilot transmissions.

\section{Frontiers of RIS Channel Modeling}
\label{sec:practical-issues}

Powerful channel estimation techniques depend on accurate, yet convenient, channel models. RISs are relatively new devices whose channel modeling brings together aspects of electromagnetic engineering, hardware constraints, and communication system concepts. Certain frontiers of RIS channel modeling are still being explored; this section outlines several issues of contemporary interest and investigation in RIS channel modeling. A summary of the contents of this section appears in Table~\ref{tab:OutstandingProblems}.

\renewcommand{\arraystretch}{1.6}
\begin{table*}[h]
\small
    \centering
    \begin{tabular}{|p{1.6 in}|p{2.2 in}|p{2.3 in}|}
        \hline
         \multicolumn{1}{|c}{\bf{Issue}} & \multicolumn{1}{|c}{\bf{Cause}} & \multicolumn{1}{|c|}{\bf{ Application or Limitation}}  \\ 
         \hline
         \hline
         Channel reciprocity & Angle dependence of RIS phase shifts & Massive MIMO channel estimation \\ 
         Mutual coupling & Reduced element spacing & Modeling \& estimation in large RIS \\
         Perfect absorption & 
         Dissipation and resonance control
         & Required in some channel estimation methods\\
         Dependence of RIS gain/phase & Metallic/Dielectric/Ohmic losses vs. phase & Passive beamforming, DFT-training\\
         RIS frequency dependence & Load/control impedance vs. frequency & Wideband communications using RIS \\
         Near field issues & Spherical vs.\ planar wave approximation & Large-RIS, indoor communication \\
         \hline
    \end{tabular}
    \caption{Frontiers of RIS channel modeling}
    \label{tab:OutstandingProblems}
\end{table*}
\subsection{Channel Reciprocity}
\begin{figure}[h]
    \centering
    \includegraphics[width=3. in]{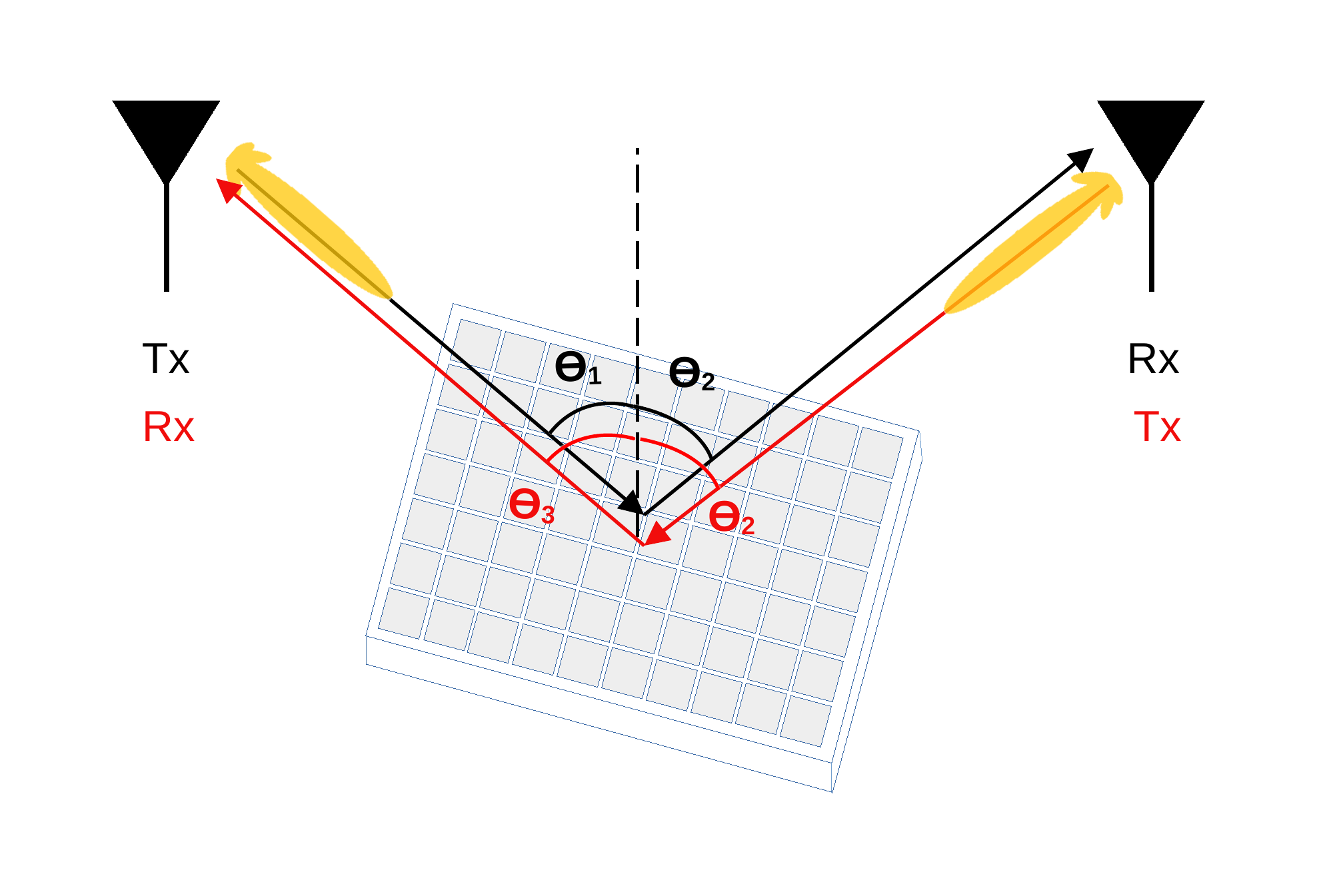}
    \vspace{-0.5 cm}
    \caption{Schematic representation of reciprocity in RIS.}
    \label{fig:Reciprocity}
\end{figure}

RIS channel estimation in multiuser settings rely on the principle of channel reciprocity for reducing the estimation and feedback overhead, since reciprocity enables the downlink precoding using uplink pilots/estimation in the time division duplex operation \cite{9400843, 9053695, zegrar2020general, 8844787}. 
Channel reciprocity holds for many boundary conditions occurring in wireless communication, including reflections from large objects as well as scattering. However, in the case of RIS-assisted systems, the literature is inconclusive (see Fig.~\ref{fig:Reciprocity}). Chen {\em et al.} \cite{9099621} based on an equivalent circuit model claims that angle reciprocity only holds for small angles with respect to normal.  In the opposite direction, Tang {\em et al.}~\cite{9690479} invokes the Rayleigh-Carson theorem to conclude that RIS enjoy reciprocity, but does not elaborate. Liang {\em et al.}~\cite{9632392} states that the angle reciprocity depends on the design of the RIS surface, and proposes a structure that achieves reciprocity for wide range of angles.

A resolution of the differences between these results, and a conclusive determination of the conditions under which RIS-induced channels are reciprocal or non-reciprocal, will be welcomed. The system model and applications for a non-reciprocal RIS may be of interest in future applications, but is in need of verifiable theory and/or experimental evidence.

\subsection{Mutual Coupling}
A common assumption in RIS modeling is that the passive elements are spaced half-wavelength apart and the mutual coupling among the elements is negligible, allowing them to be controlled individually and independently. However, in practical planar RIS structures with fixed aperture, it is desirable to increase the number of elements by reducing the inter element spacing in order to increase the directivity of the reflected waves and thereby improve the received power. Reducing the inter-element spacing results in dependency/connection of the impedances of the neighboring elements that is non-negligible, having its effect on the channel model, estimation, and the design of reflection coefficients \cite{9140329,9145091}. Gradoni and Di Renzo \cite{9319694} have proposed an electromagnetic compliant end-to-end channel model for RIS-aided communication that accounts for mutual impedance among RIS elements, while also including the effects of antenna elements at the transmitter and receiver. This impedance-based communication model is utilized in \cite{9360851} to maximize the end-to-end received power by optimizing the RIS tunable load impedances in SISO system, which is further generalized in \cite{9525465} for MIMO systems. In the end-to-end channel modeling of the aforementioned references, the statistical components of the Tx-RIS and RIS-Rx channels are intertwined with the circuit model parameters, which is not desirable from the signal processing/system design perspective. A more interesting and useful model is one that retains the factored form $\TxRisChannel\RisPhaseMatrix\RisRxChannel$, separating the statistical part of the Tx-RIS and RIS-Rx channels, but incorporating the effects of mutual impedances and  coupling into the RIS matrix $\RisPhaseMatrix$, making it potentially non-diagonal and function of circuit model parameters. Obtaining such a factored model and the associated estimation technique is a potential direction for future research.   

\subsection{Perfect Absorption/Reflection at RIS}
Several channel estimation methods, such as those based on  matrix completion and channel decomposition, depend on the ability to completely deactivate some RIS elements. Some methods depend on separately estimating the direct channel between the transmitter and the receiver, by eliminating the effect of RIS reflections, which requires deactivating all the elements. This requires the incident energy to either be completely absorbed by the RIS, or the incident waves to pass through the RIS. The feasibility of perfect absorption is debatable, with partial results whose applicability to communication systems remains unverified. Some works that study the electromagnetics of metasurfaces suggest that perfect absorption is possible at resonant frequencies with proper tuning of impedance~\cite{liu2019intelligent, imani2020perfect, li2020tunable, alaee2017theory}, but it is unclear if/how the proposed structures and the associated methods can be utilized in the context of RIS-aided communication. The issue of perfectly deactivating the passive elements is raised in \cite{9328501, 8879620, 9090876}, but the question of its physical feasibility remains unresolved.
%
Mishra and Johansson~\cite{8683663} assume that perfect reflection and absorption are unrealizable and incorporates two constants as implementation errors in the system model. To the best of the authors' knowledge, no currently-available study in the open literature offers an in-depth and conclusive treatment of the feasibility of perfect deactivation of RIS elements and related design issues. Also, analyzing the effect of imperfect deactivation on the accuracy of estimation methods is an important future direction for research.

\textcolor{black}{In a related direction, several works explore whether and how the phase and amplitude response of an RIS elements are related~\cite{9115725,4619755,9133266}. A few studies in reflectarrays and meta surfaces \cite{he2018high,9115725,zhu2013active,7744497} aim at designing structures that allow near-independent control of reflection amplitude and phase, however, their applicability to RIS-aided communication has not been established.}

\subsection{Frequency Dependence of RIS Elements}
RIS-assisted wideband communications \cite{9149146, 9409636, 9064547} requires RIS elements to efficiently operate throughout the frequencies of the band. In typical RIS constructions, however, the reflection coefficients are tuned by \textcolor{black}{switching on or off various reactive elements or patterns that are connected to the RIS element. The effect of these tuning devices can be modeled by an equivalent circuit whose load impedance varies with the carrier frequency. The phase shift applied to the incident wave via the tunable elements is calculated at a specific frequency, often the resonant frequency of the RIS element. Within a small deviation of this center frequency, the phase shift remains linear, but across wider frequencies of operation, the phase shift might vary nonlinearly. In that case, the array factor will vary across frequency, and the beam may not retain sharpness across the band of frequencies in which RIS must operate. Several remedies have been proposed in the neighboring literature in {\em reflectarrays}, e.g., coupling the elements to true time delay lines\cite{4589079} or by coupling multiple resonance elements\cite{1210823, 5484683, 6648436}. The applicability of these methods for RIS-aided communications is yet to be explored, and is a direction for future study.}

\subsection{Near Field Issues}
\begin{figure}
    \centering
    \includegraphics[width=3in]{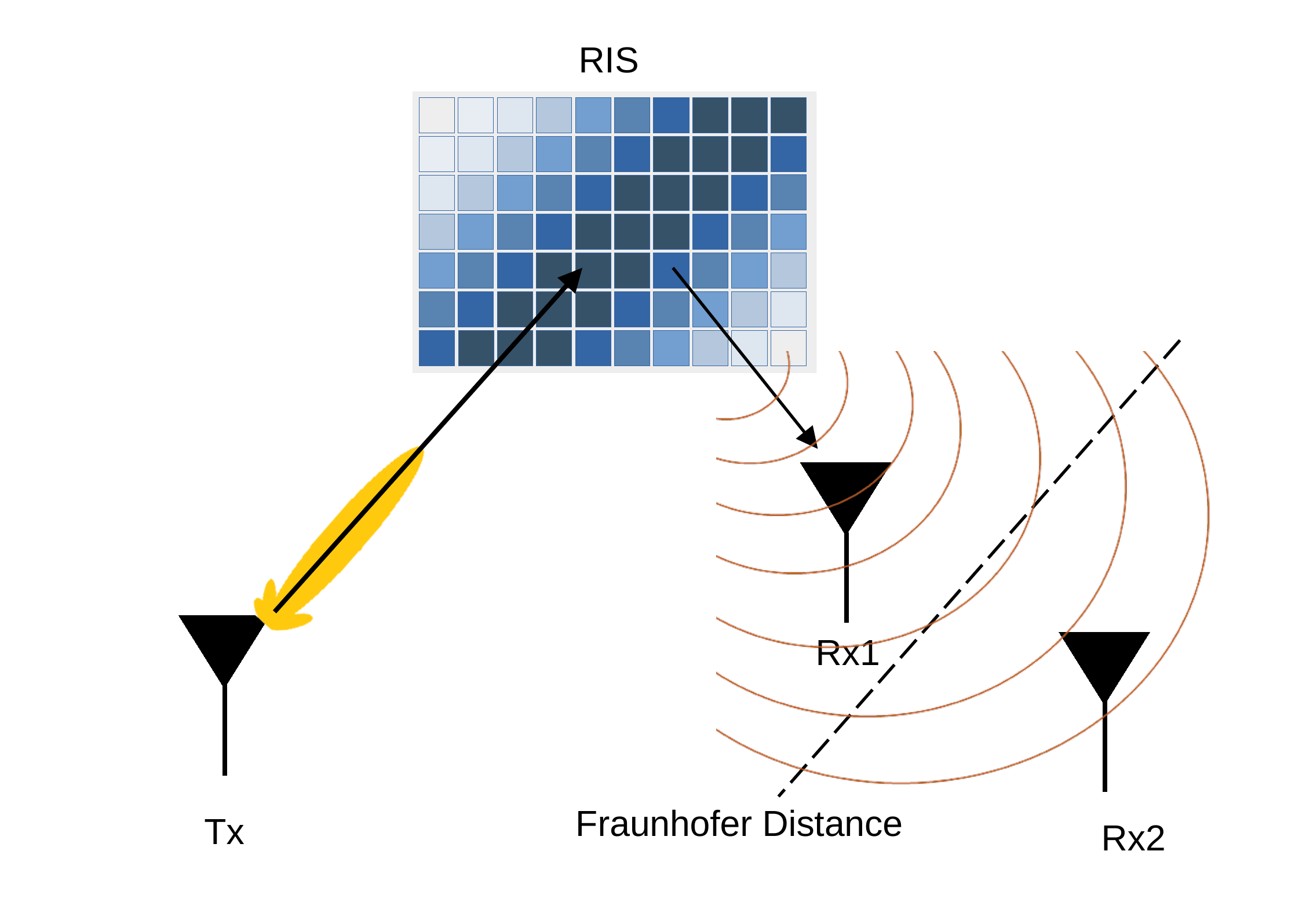}
    \vspace{0 cm}
    \caption{Illustration of near/far field issue in RIS-aided communication.}
    \label{fig:NearFarField}
\end{figure}
The transmitter/receiver is said to be in the far-field of RIS if it is at a distance greater than the Fraunhofer distance $\frac{2D_{RIS}^2}{\lambda_c}$, where $D_{RIS}$ is the largest aperture of RIS and $\lambda_c$ is the wavelength corresponding to the carrier frequency $f_c$ \cite{balanis2015antenna}. \textcolor{black}{With the far field assumption, the incident/reflected wave from the RIS can be assumed planar, which simplifies calculations (see Fig.~\ref{fig:NearFarField}). Larger RIS structures can achieve superior SNR~\cite{8811733,8796365}, but if the transmitter and receiver locations are fixed, a sufficiently large RIS array will violate the far field assumption~\cite{9184098}.} 
The near-field scenario arises, for example, when a large RIS is mounted on a large portion of the facade of a building for servicing users in the street. 
\textcolor{black}{Without the far-field assumption, the incident waves at different elements will have unequal angular directions and polarization.} The modeling of RIS in the near field scenario is considered in~\cite{9206044, 9133126, 9569465}.
The work in \cite{8319526} accounts for the difference in the effective area of the elements from different observation angles close to the RIS. Studying suitable channel models and associated estimation techniques for RIS-aided near field communication is a potential direction for future research.

\section{Conclusion}
\label{sec:conclusions}
This paper provides a comprehensive exposition of the channel estimation techniques for RIS-aided systems, ranging from classical least squares/MMSE methods to machine learning methods. RIS is often employed with many reflective elements leading to a channel gain with a huge number of parameters, therefore the estimation of link gains is a signature challenge of RIS-induced communications. This paper explores the utility of RIS channel structure for reducing the estimation overhead, including the slow variation of the base-station-to-RIS channel, sparsity of the mmWave channels, and the spatial correlations among the channels of neighboring RIS elements and neighboring users. Open problems in the broader area of RIS channel estimation were highlighted.

\bibliographystyle{IEEEtran}
\bibliography{RIS_ieee}

\end{document}